\documentclass[twocolumn,superscriptaddress,preprintnumbers,amsmath,amssymb,prc,nofootinbib]{revtex4-2}

\usepackage{graphicx}
\usepackage{dcolumn}
\usepackage{bm}
\usepackage{ifpdf}
\usepackage{epstopdf}
\usepackage{slashed}
\usepackage{amsfonts}
\usepackage{mathrsfs}
\usepackage{amssymb}
\usepackage{multirow}
\usepackage{CJK}
\usepackage{xcolor}
\usepackage{ulem}

\def\lesssim{\ \raise.3ex\hbox{$<$}\kern-0.8em\lower.7ex\hbox{$\sim$}\ }
\def\gesim{\ \raise.3ex\hbox{$>$}\kern-0.8em\lower.7ex\hbox{$\sim$}\ }

\begin{document}
\begin{CJK}{UTF8}{ipxm}
\preprint{RIKEN-iTHEMS-Report-23}

\title{BCS-BCS crossover between atomic and molecular superfluids \\
in a Bose-Fermi mixture}

\author{Yixin Guo (郭一昕)}
\email{guoyixin1997@g.ecc.u-tokyo.ac.jp}
\affiliation{Department of Physics, Graduate School of Science, The University of Tokyo, Tokyo 113-0033, Japan}
\affiliation{RIKEN iTHEMS, Wako 351-0198, Japan}

\author{Hiroyuki Tajima (田島裕之)}
\email{hiroyuki.tajima@tnp.phys.s.u-tokyo.ac.jp}
\affiliation{Department of Physics, Graduate School of Science, The University of Tokyo, Tokyo 113-0033, Japan}

\author{Tetsuo Hatsuda (初田哲男)}
\email{thatsuda@riken.jp}
\affiliation{RIKEN iTHEMS, Wako 351-0198, Japan}

\author{Haozhao Liang (梁豪兆)}
\email{haozhao.liang@phys.s.u-tokyo.ac.jp}
\affiliation{Department of Physics, Graduate School of Science, The University of Tokyo, Tokyo 113-0033, Japan}

\date{\today}

\begin{abstract}
We examine theoretically a continuity between atomic and molecular Fermi superfluids in a Bose-Fermi mixture near the Feshbach resonance.
{Considering a two-channel model with Fermi atoms $f$, Bose atoms $b$ and the closed-channel molecules $F$, we construct a mean-field framework based on the perturbative expansion of the $b$-$f$-$F$ Feshbach coupling. }
The resulting effective Hamiltonian not only exhibits the continuity between atom-atom to molecule-molecule Cooper pairings but also becomes equivalent to the two-band-superconductor model with Suhl-Matthias-Walker-type pair-exchange coupling. 
We demonstrate how these atomic and molecular Fermi superfluids coexist within the two-band-like superfluid theory.
{The pair-exchange coupling $ff \leftrightarrow FF$ and resulting superfluid gaps
$\langle ff \rangle$ and $\langle FF \rangle$}
are found to be strongly enhanced near the Feshbach resonance due to the interplay between the infrared singularity of Bogoliubov phonons and their Landau damping arising from the coupling with fermions.
The pair-exchange coupling can be probed via the observation of the intrinsic Josephson effect between atomic and molecular superfluids.
\end{abstract}

\maketitle

\section{Introduction}\label{sec:I}

Quantum simulations with ultracold atoms have opened a new frontier to investigate exotic matter in laboratories. 
One of the most illuminating examples is the experimental realization of the crossover between  the Bardeen-Cooper-Schrieffer (BCS) superfluid and the molecular Bose-Einstein condensate (BEC) in the Fermi-Fermi mixture~\cite{ketterle2008NuovoCimento}.
Furthermore, degenerate Bose-Fermi mixtures such as $^{23}$Na-$^6$Li~\cite{Hadzibabic2002PhysRevLett.88.160401}, $^{87}$Rb-$^{40}$K~\cite{Goldwin2004PhysRevA.70.021601,Marco2019Science.363.853}, $^{173}$Yb-$^{174}$Yb~\cite{Fukuhara2009PhysRevA.79.021601}, $^{84}$Sr-$^{87}$Sr~\cite{Tey2010PhysRevA.82.011608},  $^{41}$K-$^{40}$K~\cite{Wu2011PhysRevA.84.011601}, and $^{7}$Li-$^{6}$Li~\cite{Ferrier2014Science.345.1035,ikemachi2016all} have been realized and the Feshbach resonance in a $^{23}$Na--$^{40}$K mixture has also been observed \cite{Park2012PhysRevA.85.051602}.
The strong interaction between a fermion and a boson in such systems may lead to unique quantum phenomena such as the quantum phase transition from a polaronic to a molecular phase ~\cite{duda2021transition} and the realization of  long-lived Bose-Fermi Feshbach molecules with the $p$-wave interaction in a $^{23}$Na-$^{40}$K mixture~\cite{duda2022longlived}. 
Moreover, the field-linked resonance~\cite{chen2023field}, which enables us to tune the molecule-molecule interaction, has been realized in a $^{23}$Na-$^{40}$K mixture towards the realization of molecular superfluid~\cite{BARANOV200871}.

\begin{figure}
    \centering
\includegraphics[width=8.5cm]{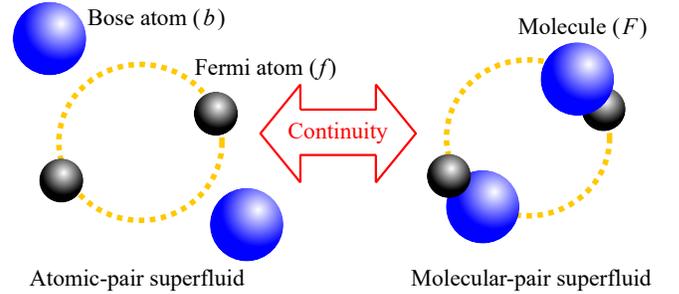}
    \caption{Continuity between atomic-pair and molecular-pair BCS superfluids in the presence of a background BEC.}
    \label{fig:continuity}
\end{figure}

Theoretically, the $p$-wave pairing of spin-polarized fermions caused by the induced interaction in a BEC background is anticipated at low temperature~\cite{Kinnunen2018PhysRevLett.121.253402}.
Then it is an interesting question whether such $p$-wave atomic Fermi superfluidity changes to the molecular Fermi superfluidity with an increasing attraction between the boson and the fermion.
A fascinating scenario is that in which the two phases show a BCS-BCS crossover between atomic and molecular superfluids as illustrated in Fig.~\ref{fig:continuity}.
Incidentally, such a crossover between the $p$-wave superfluids has been discussed in the context of the high-density quantum chromodynamics (QCD) matter realized inside neutron stars~\cite{Fujimoto:2019sxg}, in which the $p$-wave down-quark pairing in the Bose condensate of bosonic diquarks is smoothly connected to the $p$-wave pairing of neutrons (a composite fermion composed of quarks and diquarks).\footnote{In high-density QCD, such a phenomenon is generally called quark-hadron crossover or quark-hadron continuity~\cite{Schafer1999PhysRevLett.82.3956,Hatsuda2006PhysRevLett.97.122001} and its quantum simulation using the Bose-Fermi mixture of ultracold atoms was discussed in Ref.~\cite{Maeda2009PhysRevLett.103.085301}.} 
In this sense, the BCS-BCS crossover of Fermi superfluids in a Bose-Fermi mixture can help the further understanding of extremely dense matter in compact stars~\cite{Baym2018Rep.Prog.Phys.81.056902}.

In the present study, we explore a theoretical possibility of the continuity between atomic and molecular Fermi superfluid states in a Bose-Fermi mixture near the Feshbach resonance
{by considering a two-channel model with Fermi atoms $f$, Bose atoms $b$, and the closed-channel molecules $F$. }
We perform a perturbative expansion with respect to the Feshbach coupling $g$, which can be justified in the case of narrow Feshbach resonance where the small $g$ is characterized by the largely negative effective range~\cite{Chin2010RevModPhys.82.1225}.
{
In this regard, our work involves new scattering processes ($ff \leftrightarrow FF$ and $fF \leftrightarrow Ff$)
even within $O(g^2)$, which are absent in the previous works for the single-channel model with the contact-type interactions 
~\cite{Viverit2000Phys.Rev.A61.053605,Viverit2002Phys.Rev.A66.063604,Efremov2002Phys.Rev.B65.134519,Fratini2010Phys.Rev.A81.051605,Guidini2015Phys.Rev.A91.023603}.}
The effective interaction induced by the anomalous propagator of Bogoliubov phonons causes the pair-exchange interaction between two-atomic and two-molecular pairing states.
This nontrivial interaction is similar to the pair scattering extensively discussed in multiband superconductors, called the Suhl-Mattias-Walker (SMW) or Suhl-Kondo mechanism~\cite{Suhl1959PhysRevLett.3.552,Kondo196310.1143/PTP.29.1}.
We demonstrate a continuity between two-atom and two-molecule pairing condensates by solving the two-band-like mean-field theory.

While we consider the spinless fermions with the $p$-wave pairing relevant to the recent experiment in Ref~\cite{duda2021transition},
the present formalism can be easily extended to the spin-$\frac{1}{2}$ case with the $s$-wave pairing~\cite{Maeda2009PhysRevLett.103.085301}.
Although we use a simplified form of the interactions to demonstrate the continuity,
the present model can also be applied to condensed-matter systems such as multi-band $p$-wave superconductors for the purpose of qualitative understanding of pairing properties as well as an analogy with ultracold atoms and dense QCD. 
In addition, the present study suggests an alternative route to realize the two-band Fermi superfluidity with the incipient flat band~\cite{aoki2020theoretical} in a mass-imbalanced Bose-Fermi mixture.

This paper is organized as follows.
In Sec.~\ref{sec:2} we show the theoretical model of the atomic Bose-Fermi mixture and present the effective Hamiltonian based on the perturbation theory.
In Sec.~\ref{sec:3} the numerical results are presented for the purpose of demonstration of the continuity picture in this system.
For the numerical calculations, {we consider the mass ratio
between the molecule and the fermionic atom} $m_F/m_f=(40+23)/40=1.575$, which is relevant to the recent experiments of the $^{23}$Na-$^{40}$K mixture~\cite{duda2021transition,duda2022longlived}.
In Sec.~\ref{sec:4} we discuss the potential experimental observation of the continuity between atomic and molecular superfluids, such as the excitation gap and the Josephson effect to see the direct consequence of the pair-exchange coupling.
We summarize the findings in Sec.~\ref{sec:5}. 

\section{Spinless Bose-Fermi mixture near the narrow Feshbach resonance}\label{sec:2}

We consider the spinless Bose-Fermi mixture near the single Feshbach resonance, which is directly relevant to the recent experiment in Ref.~\cite{duda2021transition}.
The two-channel Hamiltonian of the Bose-Fermi mixture is given by
\begin{align}
\label{eq:H}
    H=H_{\rm Bose}+H_{\rm Fermi}+V_{Fbf},
\end{align}
where $H_{\rm Bose}=K_{b}+V_{bb}$ and $H_{\rm Fermi}=K_{f}+V_{ff}+K_{F}+V_{FF}$ are the bosonic and fermionic parts of the Hamiltonian, respectively. 
The kinetic terms read
\begin{subequations}
\begin{align}
K_{b}=\,&
\sum_{\bm{p}} \varepsilon_{\bm{p}, b} 
b_{\bm{p}}^{\dagger} b_{\bm{p}}, \\
K_{f}=\,&
\sum_{\bm{p}}
\varepsilon_{\bm{p}, f} f_{\bm{p}}^{\dagger} f_{\bm{p}}, \\
K_{F}=\,&
\sum_{\bm{p}}\varepsilon_{\bm{p},F}F_{\bm{p}}^\dag F_{\bm{p}}, 
\end{align}
\end{subequations}
where $b^\dagger$ and $f^\dagger$ create a boson and a fermion in the open channel, respectively, while $F^{\dagger}$ denotes a creation operator of a closed-channel molecular fermion.
The single-particle energies are given by
\begin{align}
    \varepsilon_{\bm{p},i}=\frac{p^2}{2m_i}-\mu_{i} \ \ \ (i=b, f, F).
\end{align}
Here $\mu_{i}$ are the chemical potentials satisfying the relation
\begin{align}
 \mu_{F}=\mu_{f}+\mu_b-\nu_{F} \equiv \mu_f -\tilde{\nu}_F,
\end{align}
with $\nu_{F}$ the closed-channel molecular energy and $\tilde{\nu}_{F}$ the renormalized energy.
The mass of the molecular fermion is given by $m_{F}=m_{f}+m_b$. 

We consider the background fermion-fermion interactions given by
\begin{align}
    V_{ff}=&\,\frac{1}{2}\sum_{\bm{k},\bm{k}',\bm{q}}
    {U_{ff}(\bm{k},\bm{k}')}\cr
    &\times f_{\bm{k}+\bm{q}/2}^\dag
    f_{-\bm{k}+\bm{q}/2}^\dag
    f_{-\bm{k}'+\bm{q}/2}
    f_{\bm{k}'+\bm{q}/2}
\end{align}
and
\begin{align}
    V_{FF}=&\,\frac{1}{2}\sum_{\bm{k},\bm{k}',\bm{q}}
    {U_{FF}(\bm{k},\bm{k}')}\cr
    &\times F_{\bm{k}+\bm{q}/2}^\dag
    F_{-\bm{k}+\bm{q}/2}^\dag
    F_{-\bm{k}'+\bm{q}/2}
    F_{\bm{k}'+\bm{q}/2},
\end{align}
where the coupling strength $U_{ff(FF)}(\bm{k},\bm{k}')$ is momentum dependent because the momentum-independent (contact) $s$-wave part vanishes due to
the {Fermi statistics.}
Also, the repulsive boson-boson interaction with the coupling strength $g_{bb}$ is given by
\begin{align}
V_{bb}=\,&
\frac{1}{2} g_{bb} \sum_{\bm{P}, \bm{q}, \bm{q}'} 
b_{{\bm{P}}/{2}+\boldsymbol{q}}^{\dagger} 
b_{{\bm{P}}/{2}-\boldsymbol{q}}^{\dagger} 
b_{{\bm{P}}/{2}-\boldsymbol{q}' } 
b_{{\bm{P}}/{2}+\boldsymbol{q}' },
\end{align}
which develops a BEC in the mean-field level with {$\langle b_{\bm{0}}^\dag\rangle =\langle b_{\bm{0}}\rangle^*=\sqrt{\rho_{b}}\, e^{i\theta_{\rm BEC}}$, where 
the condensate bosonic density and its phase are denoted by $\rho_b$ and $\theta_{\rm BEC}$, respectively.
{Note that the $s$-wave interaction is sufficient to describe the low-energy boson-boson interaction in contrast to that between two identical fermions.}
The Feshbach coupling of the composite fermion $F$ (closed-channel molecule) to the boson $b$ and the fermion $f$} with the coupling strength $g$ is given by
\begin{align}
    V_{Fbf}=\sum_{\bm{P}, \bm{q}}g\left(
    F_{\bm{P}}^\dag 
    b_{{\bm{P}}/{2}-\bm{q}}
    f_{{\bm{P}}/{2}+\bm{q}}+{\rm H.c.}\right).
\end{align}
The scattering length $a_{bf}$ and the effective range $r_{bf}$ between $b$ and $f$ at low energies are related to $g$ and $\nu_F$ as~\cite{OHASHI2020103739}
\begin{align}
\label{eq:abf}
    \frac{2\pi a_{bf}}{m_{r}}&=\left[-\frac{\nu_F}{g^2}
    +\sum_{\bm{q}}\frac{1}{\varepsilon_{\bm{q},b}+\varepsilon_{\bm{q},f}}\right]^{-1}, \\
    r_{bf}&=-\frac{2\pi}{m_r^2g^2},
\end{align}
where $m_r=\frac{m_fm_b}{m_f+m_b}$ is the reduced mass.

{
Although $a_{bf}$ in Eq.~\eqref{eq:abf} approaches zero in the limit $g\rightarrow 0$, 
 the background $b$-$f$ scattering which may be characterized by the contact interaction ~\cite{Kinnunen2018PhysRevLett.121.253402}
 makes $a_{bf}$ finite in the limit ~\cite{OHASHI2020103739}.
 Under the presence of such background scattering, 
the $p$-wave Fermi superfluids (the $ff$ pairing and the $FF$ pairing) would be further enhanced, in particular, in the regime away from the resonance.
 However,  we will not consider the effect below, since our main focus is on the qualitative feature of the resonant pair-exchange coupling between $ff$ and $FF$.}

{
In the following subsections we first derive the effective interactions in Sec.~\ref{sec:IIA} and evaluate the boson self-energy associated with the Landau damping in Sec.~\ref{sec:IIB}.
Then the effective Hamiltonian is presented in Sec.~\ref{sec:IIC}.
In Sec.~\ref{sec:IID} we demonstrate the BCS-BCS crossover between the atomic ($ff$) superfluid and the molecular ($FF$) superfluid
 within the mean-field theory.}

\subsection{Effective interactions}\label{sec:IIA}

\begin{figure}[t]
    \centering
    \includegraphics[width=8cm]{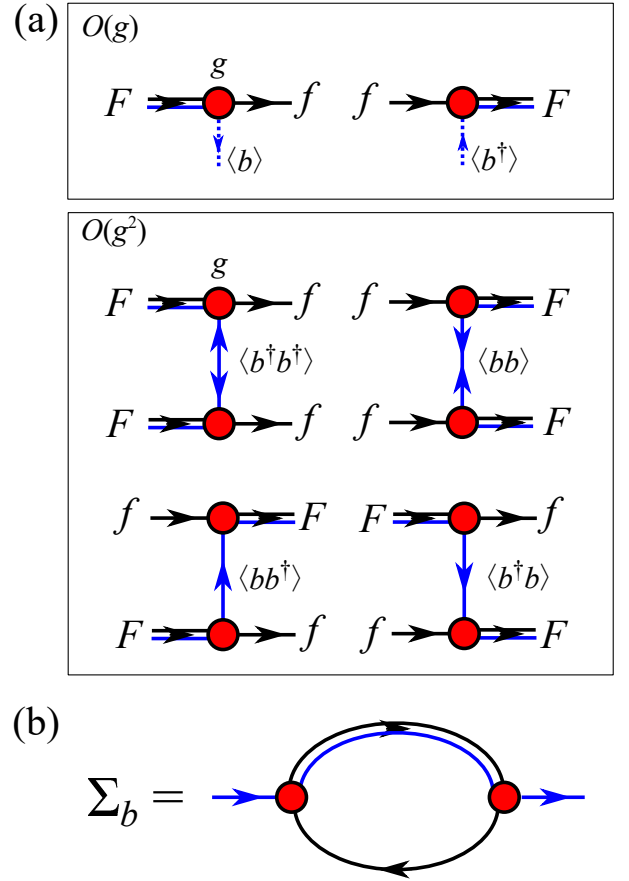}
    \caption{{(a)} Effective couplings between a fermion and a composite fermion at the leading and the next-to-leading orders of the Feshbach coupling $g$.
    {(b) Bosonic self-energy $\Sigma_b$ for the Landau damping.}
    }
    \label{fig:BF2ch}
\end{figure}

We consider the small-$g$ case, which corresponds to the narrow Feshbach resonance with the large and negative $r_{bf}$.
Under the BEC background, the one-body mixing of $O(g)$ reads  [Fig.~\ref{fig:BF2ch}(a)]
\begin{align}
    V_{\rm M}=g\sqrt{\rho_b}\sum_{\bm{P}}\left(F_{\bm{P}}^\dag f_{\bm{P}}+f_{\bm{P}}^{\dag}F_{\bm{P}}\right).
        \label{eq:VM}
\end{align}
Such a term is similar to the Rabi coupling in ultracold atoms~\cite{Goldman2014} as well as to  the interband impurity in two-band superconductors~\cite{Chow1968PhysRev.172.467}.

{In the following, the superfluid-phonon exchange in the cold-atom system plays a crucial role.}
As shown in  Fig.~\ref{fig:BF2ch}(a), the next-to-leading order terms of $O(g^2)$ are given by the SMW-type pair-exchange interaction $V_{\rm SMW}$ and the phonon-mediated (PM)-type interaction $V_{\rm PM}$ as
\begin{align}
    V_{\rm SMW}=&\frac{1}{2}\sum_{\bm{k},\bm{k}',\bm{P}}
    U_{\rm SMW}(\bm{q},\omega) \cr
    &\ \ \ \ \times f_{\bm{k}+{\bm{P}}/{2}}^\dag f_{-\bm{k}+{\bm{P}}/{2}}^\dag
    F_{-\bm{k}'+{\bm{P}}/{2}} F_{\bm{k}'+{\bm{P}}/{2}} +{\rm H.c.}, \\
    V_{\rm PhM}
    =&\frac{1}{2}\sum_{\bm{k},\bm{k}',\bm{P}}
    U_{\rm PM}(\bm{q},\omega) \cr
    &\ \ \ \ \times  F_{\bm{k}+{\bm{P}}/{2}}^\dag F_{-\bm{k}'+{\bm{P}}/{2}}
    f_{-\bm{k}+{\bm{P}}/{2}}^\dag f_{\bm{k}'+{\bm{P}}/{2}} +{\rm H.c.},
\end{align}
where $\bm{q}$ and $\omega$ are the momentum and energy transfers between fermions, respectively.
{While $\omega$ depends on $\bm{k}$, $\bm{k}'$, and $\bm{P}$ in principle, we will approximate it later by a typical  energy scale associated with the Fermi energy [Eqs.~(\ref{eq:bar}) and (\ref{eq:q0})].}
{The effective interactions $U_{\rm SMW}$ and $U_{\rm PM}$ can be expressed in terms of anomalous and normal parts of the bosonic Green's function
[$D_{12}(\bm{q},\omega)$ and $D_{11}(\bm{q},\omega) $, respectively] as}
\begin{widetext}
\begin{align}
\label{eq:U-SMW}
    U_{\rm SMW}(\bm{q},\omega)&=g^2
    D_{12}(\bm{q},\omega) 
    =g^2g_{bb}\rho_b e^{2i\theta_{\rm BEC}}
    \frac{\omega^2-E_{\bm{q},b}^2+\Gamma^2(\bm{q},\omega)-2i\omega\Gamma(\bm{q},\omega)}{[(\omega-E_{\bm{q},b})^2+\Gamma^2(\bm{q},\omega)][(\omega+E_{\bm{q},b})^2+\Gamma^2(\bm{q},\omega)]}, \\
\label{eq:U-PhM}
    U_{\rm PhM}(\bm{q},\omega)&=g^2
    D_{11}(\bm{q},\omega)
    =g^2\frac{    [\omega+\varepsilon_{\bm{q}}+g_{bb}n_0+i\Gamma(\bm{q},\omega)][\omega-E_{\bm{q},b}-i\Gamma(\bm{q},\omega)][\omega+E_{\bm{q},b}-i\Gamma(\bm{q},\omega)]
}
    {[(\omega-E_{\bm{q},b})^2+\Gamma^2(\bm{q},\omega)][(\omega+E_{\bm{q},b})^2+\Gamma^2(\bm{q},\omega)]}.
    \end{align}
\end{widetext}
Here, the imaginary part of the boson self-energy due to Landau damping is denoted as  $\Gamma(\bm{q},\omega)= - {\rm Im} \Sigma_b(\bm{q},\omega) $, while the Bogoliubov phonon dispersion reads
\begin{align}
E_{\bm{q},b}=\sqrt{\epsilon_{\bm{q},b}(\epsilon_{\bm{q},b}+2g_{bb}\rho_b)},
\end{align}
with $\epsilon_{\bm{q},b}={q^2}/({2m_b})$.
Note that the boson chemical potential is given by $\mu_b=g_{bb}\rho_b+\Sigma_0$ with $\Sigma_0$ a constant part of ${\rm Re} \Sigma_b(\bm{q},\omega)$, when the gapless condition for the phonon excitation is imposed. 
{The above expressions of $U_{\rm SMW}(\bm{q},\omega)$ for the precess $ff \leftrightarrow FF$ 
and $U_{\rm PM}(\bm{q},\omega)$ for the precess $fF \leftrightarrow Ff$  represent the interaction amplitudes involving a
 one-phonon propagator in the lowest order of the Feshbach coupling $g$ as shown in Fig.~\ref{fig:BF2ch}(a).
 Since $V_{\rm SMW}$ ($V_{\rm PM}$) contains only the anomalous (normal)  phonon propagator, an infrared singularity arises at low momentum: Such a singularity is tamed dynamically by the
Landau damping described by the phonon self-energy in Fig.~\ref{fig:BF2ch}(b).
 This mechanism  is in contrast to the $ff \leftrightarrow ff$ scattering diagram in the single-channel model~\cite{CamachoGuardian2018Phys.Rev.Lett.121.013401,CamachoGuardian2018Phys.Rev.X.8.031042}, where 
 both the normal and anomalous phonon propagators contribute simultaneously in the one-phonon exchange diagram and the infrared singularities cancel each other.}

We find that $V_{\rm SMW}$ originating from the anomalous boson propagator $D_{12}$ is similar to the  pair-exchange Hamiltonian in the context of multiband superconductors~\cite{Suhl1959PhysRevLett.3.552,Kondo196310.1143/PTP.29.1}.
In particular, the present interaction 
induces the $p$-wave $FF$ and $ff$ pairings and describes the transition between these pairs simultaneously. 
Another effective interaction $V_{\rm PM}$ originating from the normal boson propagator $D_{11}$ is similar to the phonon-mediated  electron-electron interaction in  conventional superconductors and provides an attraction in the mass-imbalanced $Ff$ pair.

For later purposes, we define the Fermi momenta and Fermi energies of $f$ and $F$ as
\begin{align}\label{eq17}
k_i = {(6\pi^2\rho_i)^{{1}/{3}}}, \ \  E_i = \frac{k_i^2}{2m_i}, \ \ \  (i=f, F), 
\end{align}
with $\rho_i$ the number density. 
We introduce the total fermion number density $\rho_0$ together with a reference momentum $k_0$ and a reference energy $E_0$ as  
\begin{align}
\rho_0 & = \rho_f + \rho_F, \\
\label{eq:k0E0}
k_0  & = {(6\pi^2\rho_0)^{{1}/{3}}}, \ \  E_0 = \frac{k_0^2}{2m_0}, 
\end{align}
with $m_0 \equiv \frac{m_fm_F}{m_f+m_F}$.

We now consider the situation where $F$ is heavier than $f$ due to the boson mass $m_b$ (i.e., $m_F=m_f+m_b > m_f$) and is located near the resonance region ($\mu_F=\mu_f + \mu_b - \nu_F = 0$). 
The mass imbalance leads to the Fermi-surface mismatch between $F$ and $f$ (i.e., $E_f\neq E_F$) even for the population-balanced case and suppresses the BCS-type $Ff$ pairing~\cite{GUBBELS2013255}.
Moreover, in the most general case with the population imbalance, where there is a Fermi-momentum mismatch between $f$ and $F$, (i.e., $k_f\neq k_F$), the formation of the BCS-type $Ff$ pairing due to $V_{\rm PM}$ is suppressed further. 
Therefore, $V_{\rm SMW}$ is expected to provide a dominant contribution leading to the $p$-wave $FF$ and $ff$ pairings in the perturbative weak-coupling regime. 
Then the resonant scattering through the SMW-type interaction enhances both the $FF$ and $ff$ pairing gaps, analogous to the case of the two-band superconductors~\cite{Ochi2022PhysRevResearch.4.013032}.
Thus, we focus only on $V_{\rm SMW}$ in the following analysis and replace its arguments by the characteristic momentum and energy transfer as 
\begin{align}
\label{eq:bar}
U_{\rm SMW}(\bm{q},\omega)\rightarrow U_{\rm SMW}(|\bm{q}|= \bar{q},\omega=\bar{\omega}),
\end{align}
with
\begin{align}
\label{eq:q0}
\bar{q}=\sqrt{k_f^2+k_F^2-2k_fk_F\cos\theta_{\bm{k}\bm{k}'}}, \ \ \ 
\bar{\omega}=E_f-E_F.
\end{align}
Here, $\theta_{\bm{k}\bm{k}'}$ is the angle between $\bm{k}_f$ and $\bm{k}_F$, which is essential to obtain the effective $p$-wave interaction.
We note that the Landau damping factor $\Gamma(\bar{q},\bar{\omega})$ is crucial to tame the infrared divergence of $\left. U_{\rm SMW}(\bar{q},\bar{\omega})\right|_{\Gamma=0}$ at $\bar{\omega}=\pm E_{\bar{q},b}$.

\subsection{Effect of Landau damping}\label{sec:IIB}

Let us evaluate the Landau damping factor originating from  the coupling between the Bogoliubov phonon with the fermions through the scattering processes, phonon+$f$ $\rightarrow F$ and $F\rightarrow $ phonon+$f$. 
{As shown diagrammatically in Fig.~\ref{fig:BF2ch}(b), this process can be described by the one-loop diagram.} 
The imaginary part of the time-ordered phonon self-energy~\cite{fetter2012quantum} reads
\begin{align}
\label{eq:sigma_b}
    {\rm Im}\Sigma_{b}(\bm{q},\omega)
    =&-\pi g^2 {\rm sgn}(\omega)\sum_{\bm{p}}
    \delta(\omega-\varepsilon_{\bm{p}+\bm{q},F}+\varepsilon_{\bm{p},f})\cr
    & \ \ \ \ \ \ \ \ \times[n(\varepsilon_{\bm{p},f})-n(\varepsilon_{\bm{p}+\bm{q},F})],
\end{align}
where $n(x)$ is the Fermi distribution function.
At $T=0$, Eq.~\eqref{eq:sigma_b} can be evaluated as 
\begin{widetext}
\begin{align}
    {\rm Im}\Sigma_{b}(\bm{q},\omega)
    =-\frac{g^2}{8\pi q} \theta\left(q^2-2m_b\Omega\right){\rm sgn}(\omega)
    &\left[m_Fk_{f}^2\theta(k_{f}-Q_{+})+m_FQ_{+}^2\theta(Q_{+}-k_{f})-m_FQ_{-}^2\right.\cr
    -
    &\left.
    m_fk_{F}^2\theta(k_{F}-P_{+})-m_fP_{+}^2\theta(P_{+}-k_{F})+m_fP_{-}^2
    \right],
\end{align}
with $\Omega=\omega-E_{f}+E_{F}$ and 
\end{widetext}
\begin{align}
 Q_{\pm}&=\frac{1}{2m_F\zeta}\left|\sqrt{\frac{m_F}{m_f}(q^2-2m_b\Omega)}
    \pm q\right|, \\
P_{\pm} & =\frac{1}{2m_f\zeta}\left|\sqrt{\frac{m_f}{m_F}(q^2-2m_b\Omega)}\pm q\right|, \\
\zeta & = \frac{1}{2} \left( \frac{1}{m_f}- \frac{1}{m_F} \right).
\end{align}
Substituting $|\bm{q}|=\bar{q} $ and $\omega=\bar{\omega}$ (i.e., $\Omega=0$)
into ${\rm Im}\Sigma_b(\bm{q},\omega)$, we obtain
\begin{align}
\label{eq:gamma}
\Gamma(\bar{q},\bar{\omega}) & = \alpha\ \frac{|\bar{\omega}|}{\bar{q}}, \\ 
 \alpha & =\frac{m_fm_Fg^2}{4\pi}.
\end{align}
We note that Eq.~(\ref{eq:gamma}) is valid for a small momentum transfer $q<{\rm min}\left(\frac{\sqrt{m_F}-\sqrt{m_f}}{\sqrt{m_f}}k_f, \frac{\sqrt{m_F}-\sqrt{m_f}}{\sqrt{m_F}}k_F \right)$.
To simplify the following analysis, we consider the case that the momentum scale associated with the bosonic condensate is much larger than $k_f$ and $k_F$,
\begin{align}
k_b=\sqrt{4m_bg_{bb}\rho_b} \gg k_{f,F}.
\label{eq:k_b}
\end{align}
In this case, we have { ${\bar{\omega}}/{(g_{bb}\rho_b)}=O(k_i k_j/k_b^2)$ and ${E_{\bar{q},b}}/{(g_{bb}\rho_b)}=v_b\bar{q}+O(k_i k_j/k_b^2)$,} where $i$ and $j$ take either $f$ or $F$.
Then Eq.~(\ref{eq:U-SMW}) becomes 
\begin{align}
\label{eq:approx}
    U_{\rm SMW}(\bar{q},\bar{\omega})
    &\simeq g^2g_{bb}\rho_be^{2i\theta_{\rm BEC}}\frac{\alpha^2\frac{\bar{\omega}^2}{\bar{q}^2}-v_b^2\bar{q}^2}{\left(\alpha^2\frac{\bar{\omega}^2}{\bar{q}^2}+v_b^2\bar{q}^2\right)^2},
\end{align}
where $v_b=\sqrt{{g_{bb}\rho_b}/{m_b}}$ is the velocity of the Bogoliubov phonon.
{When obtaining Eq.~\eqref{eq:approx},
we use the expansion
\begin{align}
    &\frac{\bar{\omega}^2-E_{\bar{q},b}^2+\Gamma^2(\bar{q},\bar{\omega})-2i\bar{\omega}\Gamma(\bar{q},\bar{\omega})}{(g_{bb}\rho_b)^2}\cr
    &\quad =
    \frac{\Gamma^2(\bar{q},\bar{\omega})-v_b^2\bar{q}^2}{(g_{bb}\rho_b)^2}
    +O(k_i k_j k_\ell/k_b^3),
\end{align}
for $i,j,\ell=f,F$ (for details of power counting, see Appendix~\ref{app:pc}).
}
Equation~\eqref{eq:approx} captures essential properties of the effective interaction with the Landau damping:
For a small momentum transfer, we find $U_{\rm SMW}(\bar{q}\rightarrow 0,\bar{\omega})\simeq g^2g_{bb}\rho_b\frac{\bar{q}^2}{\alpha^2 \bar{\omega}^2} < \infty$, since $\bar{\omega}\equiv \frac{k_f^2}{2m_f}\left(1-\frac{m_f}{m_F}\frac{k_F^2}{k_f^2}\right) \neq 0$;
for a small energy transfer, we find $U_{\rm SMW}(\bar{q},\bar{\omega}\rightarrow 0 )\simeq -g^2g_{bb}\rho_b\frac{1}{v_b^2\bar{q}^2} < \infty$, since $\bar{q} \neq 0$.
We note that the present Landau damping at $T=0$ is different from that in a pure bosonic system~\cite{Giorgini1998PhysRevA.57.2949}, which occurs only at nonzero temperature.

\

\subsection{Effective Hamiltonian}\label{sec:IIC}
The effective Hamiltonian for atomic fermions $f$ and the closed-channel molecule $F$ reads
\begin{align}
\label{eq:H_eff}
    H_{\rm eff} =&
    K_{\rm eff}+V_{\rm SMW}+E_{{\rm Bose}}
    {+V_{ff}+V_{FF}}
    ,
\end{align}
where 
\begin{align}
    K_{\rm eff}=\sum_{\bm{p}}
    (\begin{array}{cc}
    f_{\bm{p}}^\dag &  F_{\bm{p}}^\dag
    \end{array})
    \left(\begin{array}{cc}
    \varepsilon_{\bm{p},f} &  g\sqrt{\rho_b} \\
        g\sqrt{\rho_b}   & \varepsilon_{\bm{p},F}
    \end{array}\right)
    \left(\begin{array}{c}
    f_{\bm{p}}  \\
        F_{\bm{p}}   
    \end{array}\right)
\end{align}
is the effective one-body Hamiltonian for fermions.
The ground-state energy of condensed bosons evaluated in the mean-field approximation is 
\begin{align}
    E_{\rm Bose}=-\mu_b\rho_b+\frac{1}{2}g_{bb}\rho_b^2.
\end{align}

To explore the fermionic superfluid state, we apply the mean-field approximation as
\begin{widetext}
\begin{align}
    V_{\rm SMW}+V_{ff}+V_{FF}\rightarrow 
    &-\frac{1}{2}\sum_{\bm{k}}
    \Delta_{ff}^*(\bm{k})
    f_{-\bm{k}}f_{\bm{k}}
    -\frac{1}{2}\sum_{\bm{k}}
    \Delta_{ff}(\bm{k})
    f_{\bm{k}}^\dag f_{-\bm{k}}^\dag
    -\frac{1}{2}\sum_{\bm{k},\bm{k}'}
    \Delta_{FF}^*(\bm{k})
    F_{-\bm{k}}F_{\bm{k}}
    -\frac{1}{2}\sum_{\bm{k}}
    \Delta_{FF}(\bm{k})
    F_{\bm{k}}^\dag F_{-\bm{k}}^\dag\cr
    &-\frac{1}{2}\sum_{\bm{k},\bm{k}'}
       [U_{\rm SMW}(\bar{q},\bar{\omega})\langle F_{\bm{k}}^\dag F_{-\bm{k}}^\dag \rangle\langle f_{-\bm{k}'}f_{\bm{k}'}\rangle
       +
       U_{\rm SMW}^*(\bar{q},\bar{\omega})
       \langle f_{\bm{k}}^\dag f_{-\bm{k}}^\dag \rangle\langle F_{-\bm{k}'}F_{\bm{k}'}\rangle]\cr
       &-\frac{1}{2}
       \sum_{\bm{k},\bm{k}'}
       \left[U_{ff}(\bm{k},\bm{k}')
       \langle f_{\bm{k}}^\dag f_{-\bm{k}}^\dag\rangle
    \langle f_{-\bm{k}'} f_{\bm{k}'}\rangle
    +U_{FF}(\bm{k},\bm{k}')
           \langle F_{\bm{k}}^\dag F_{-\bm{k}}^\dag\rangle
    \langle F_{-\bm{k}'} F_{\bm{k}'}\rangle
       \right],
\end{align}
where we introduce the pairing order parameters
\begin{align}
    \Delta_{ff}(\bm{k})
    &=-\sum_{\bm{k}'}
    \left[U_{\rm SMW}^*(\bar{q},\bar{\omega})\langle F_{-\bm{k}'} F_{\bm{k}'}\rangle
    +U_{ff}(\bm{k},\bm{k}')\langle f_{-\bm{k}'}f_{\bm{k}'}\rangle\right], \\
    \Delta_{FF}(\bm{k})
    &=-\sum_{\bm{k}'}
    \left[U_{\rm SMW}(\bar{q},\bar{\omega})\langle f_{-\bm{k}'} f_{\bm{k}'}\rangle
    +U_{FF}(\bm{k},\bm{k}')\langle F_{-\bm{k}'}F_{\bm{k}'}\rangle\right].
\end{align}
\end{widetext}

We perform the partial-wave decomposition as~\cite{Ho2005PhysRevLett.94.090402}
\begin{align}
    U_{\rm SMW}(\bar{q},\bar{\omega})&
    =4\pi \sum_{\ell=0}^\infty
    \sum_{m=-\ell}^{m=\ell}
    U_{\ell m}(k_f,k_F)
    {Y_{\ell m}(\hat{\bm{k}}) [Y_{\ell m}(\hat{\bm{k}}')]^*,
    }
\end{align}
where $\hat{\bm{k}}=\bm{k}/k$ and $k=|\bm{k}|$.
We note that $ U_{\rm SMW}(\bar{q},\bar{\omega})$ has the $\theta_{\bm{k}\bm{k}'}$-dependence via Eq.~(\ref{eq:q0}).
Because of the Pauli principle, the $s$-wave pairings ($ff$ and $FF$) are prohibited.
On the other hand, the $p$-wave pairings are induced by the effective interaction
{$U_{\ell=1 m}(k_f,k_F)$, whose explicit form (Appendix~\ref{App:U1})} is given by
\begin{align}
\label{eq:U1}
    U_{1 m} (k_f,k_F) \simeq &
    -\delta_{m,0}\frac{3m_bg^2e^{2i\theta_{\rm BEC}}(k_f^2+k_F^2)}{16k_f^2k_F^2}\cr
    &\times \ln\left[\frac{\alpha^2\bar{\omega}^2+v_b^2(k_f-k_F)^4}{\alpha^2\bar{\omega}^2+v_b^2(k_f+k_F)^4}
    \right],
\end{align}
where we keep the term involving the logarithmic singularity relevant in the weak coupling.
{
In our present two-channel model, for the pair-exchange process (i.e., $ff \leftrightarrow FF$), only the anomalous part of the single-phonon propagator contributes and causes a logarithmic singularity at low momentum unless there is no Landau damping. 
As we mentioned in Sec.II.A,  this is in sharp contrast with previous studies of the single-channel model~\cite{CamachoGuardian2018Phys.Rev.Lett.121.013401,CamachoGuardian2018Phys.Rev.X.8.031042} for the $ff \rightarrow ff$ process, where both the normal and anomalous phonon propagators contribute even in the single-phonon exchange.
}

Based on Eq.~\eqref{eq:U1}, we arrive at the density-dependent pair-exchange $p$-wave interaction,
\begin{align}
\label{eq:USMW_dens}
    U_{\rm SMW}(\bm{q},\omega)&=\lambda
    kk'
{ Y_{10}(\hat{\bm{k}}) Y_{10}(\hat{\bm{k}'}),} \\
    \lambda&=4\pi\frac{U_{10}(k_f,k_F)}{k_fk_F}.
\end{align}
Similarly, we also consider the residual interactions in the same $p$-wave channel as
\begin{align}
\label{eq:uFF}
    U_{FF}(\bm{k},\bm{k}')&=\lambda_{FF}
{     kk'Y_{10}(\hat{\bm{k}}) Y_{10}(\hat{\bm{k}'}),}
    \\
    \label{eq:uff}
    U_{ff}(\bm{k},\bm{k}')&=\lambda_{ff}
{  kk'Y_{10}(\hat{\bm{k}}) Y_{10}(\hat{\bm{k}'}) ,}
\end{align}
where the coupling constants can be expressed by the $p$-wave scattering volume $v_{j=f,F}$ as 
\begin{align}
    \frac{m_j}{4\pi v_j}=\frac{1}{{\lambda}_{jj}}+\frac{m_jk_{\rm cut}^3}{24\pi^3}.
\end{align}
We introduce a momentum cutoff $k_{\rm cut}=k_b$, because the effective interaction is valid for large $k_b$ compared to the other fermionic momentum scales (i.e., $k$, $k'$, and $k_{f,F}$).
Although the cutoffs for each interaction are not necessarily equal to each other, we use a common cutoff $k_{b}$ for simplicity.
This approximation is valid in the weak-coupling regime, where the momentum near $k_{\rm F}$ is relevant.

\subsection{Mean-field energy}\label{sec:IID}

Eventually, we obtain the $4\times 4$ matrix mean-field Hamiltonian
\begin{widetext}
\begin{align}
\label{eq:H_MF}
     H_{\rm eff}^{\rm MF}
    =&\,\frac{1}{2}\sum_{\bm{p}}\Psi_{\bm{p}}^\dag
    \left(\begin{array}{cccc}
        \varepsilon_{\bm{p},f} & -\Delta_{ff}(\bm{p}) & \gamma & 0 \\
        -\Delta_{ff}^*(\bm{p}) & -\varepsilon_{\bm{p},f}  &
        0 & -\gamma \\
         \gamma& 0 & \varepsilon_{\bm{p},F} & -\Delta_{FF}(\bm{p}) \\
         0 & -\gamma & -\Delta_{FF}^*(\bm{p})  & -\varepsilon_{\bm{p},F} 
    \end{array}\right)\Psi_{\bm{p}}+\frac{1}{2}\sum_{\bm{p}}[\varepsilon_{\bm{p},f}+\varepsilon_{\bm{p},F}]\cr
    &- \frac{1}{4}|\lambda||d_{FF}||d_{ff}|
       \cos(\theta_{ff}-\theta_{FF}+2\theta_{\rm BEC})
       +\frac{1}{2}|\lambda_{ff}||d_{ff}|^2
       +\frac{1}{2}|\lambda_{FF}||d_{FF}|^2
    +E_{{\rm Bose}},
\end{align}
\end{widetext}
where $\Psi_{\bm{p}}=(f_{\bm{p}} \ f_{-\bm{p}}^\dag \ F_{\bm{p}} \ F_{-\bm{p}}^\dag )^{\rm T}$ is the four-component Nambu spinor, and $\gamma \equiv g\sqrt{\rho_b}$ is the strength of one-body mixing in Eq.~(\ref{eq:VM}).
The separability of the interaction leads to the simplified form of the pairing order parameters
\begin{align}
\label{eq:delta_ff}
    \Delta_{ff}(\bm{k})
    &=-\sum_{\bm{k}'}
  {   kk'Y_{10}(\hat{\bm{k}}) Y_{10}(\hat{\bm{k}}') }\cr
    &\qquad\times\left(\lambda^*\langle F_{-\bm{k}'}
    F_{\bm{k}'}\rangle
    +\lambda_{ff}\langle f_{-\bm{k}'}
    f_{\bm{k}'}\rangle
    \right)\cr
    &\equiv kY_{10}(\hat{\bm{k}})(\lambda^* d_{FF}+\lambda_{ff}d_{ff}),\\
    \Delta_{FF}(\bm{k})
    &=-
    \sum_{\bm{k}'}
{  kk'Y_{10}(\hat{\bm{k}}) Y_{10}(\hat{\bm{k}}')} \cr
    &\qquad\times\left(\lambda\langle f_{-\bm{k}'}
    f_{\bm{k}'}\rangle
    +\lambda_{FF}\langle F_{-\bm{k}'}
    F_{\bm{k}'}\rangle
    \right)\cr
&\equiv kY_{10}(\hat{\bm{k}})(\lambda d_{ff}+\lambda_{FF}d_{FF}).
\label{eq:delta_FF}
\end{align}

Note that the mean-field Hamiltonian $H_{\rm eff}^{\rm MF}$ does not have  global symmetry because of the finite BCS pairings, $\langle FF\rangle\neq 0$ and $\langle ff \rangle\neq 0$, similar to the case of two-band superconductors~\cite{Mihail2012PhysRevB.85.134514}.

Eigenvalues of the $4\times 4$ matrix in $H_{\rm eff}^{\rm MF}$ are
\begin{subequations}
\begin{align}
e_{1}^{(\mp)}=\,&
\mp \left[\frac{1}{2}\mathcal{A}
+
\gamma^2
-
\frac{1}{2}
\sqrt{
\left(
\mathcal{A}
+2\gamma^2
\right)^2
-4
\mathcal{B}
}
\right]^{1/2},\\
e_{2}^{(\mp)}=\,&
\mp \left[\frac{1}{2}\mathcal{A}
+
\gamma^2
+
\frac{1}{2}
\sqrt{
\left(
\mathcal{A}
+2\gamma^2
\right)^2
-4
\mathcal{B}
}
\right]^{1/2},
\end{align}
\end{subequations}
where
\begin{subequations}
\begin{align}
\mathcal{A}=\,&\varepsilon_{\bm{p},f}^2
+\varepsilon_{\bm{p},F}^2
+\left|\Delta_{ff}(\bm{p})\right|^2
+\left|\Delta_{FF}(\bm{p})\right|^2,\\
\mathcal{B}=\,&
\left(\gamma^2-\varepsilon_{\bm{p},f}\varepsilon_{\bm{p},F}\right)^2
+\varepsilon_{\bm{p},f}^2
\left|\Delta_{FF}(\bm{p})\right|^2
\nonumber\\&
+\varepsilon_{\bm{p},F}^2\left|\Delta_{ff}(\bm{p})\right|^2 
+
\left|\Delta_{FF}(\bm{p})\right|^2
\left|\Delta_{ff}(\bm{p})\right|^2
\nonumber\\&
+
2 \gamma^2
{\rm Re} \left[
\Delta^\ast_{ff}(\bm{p})
\Delta_{FF}(\bm{p})
\right].
\end{align}
\end{subequations}
Thus, we obtain the ground-state energy as
\begin{align}
\label{eq:E0}   
E_{\rm GS}=\frac{1}{2}\sum_{\bm{p}}\left( \varepsilon_{\bm{p},f}+\varepsilon_{\bm{p},F} + e_{1}^{(-)} + e_{2}^{(-)}  \right)
     +E_{\rm J} +E_{\rm Bose}.
\end{align}
The fermionic constant term given by
\begin{align}
    E_{\rm J}=&- \frac{1}{4}|\lambda||d_{FF}||d_{ff}|
       \cos(\theta_{ff}-\theta_{FF}+2\theta_{\rm BEC}) \nonumber\\ &
       +\frac{1}{2}|\lambda_{ff}||d_{ff}|^2
       +\frac{1}{2}|\lambda_{FF}||d_{FF}|^2
\end{align}
is called the Josephson coupling energy {with $d_{FF}=|d_{FF}|e^{i\theta_{FF}}$,
$d_{ff}=|d_{ff}|e^{i\theta_{ff}}$, and $\lambda=|\lambda|e^{2i \theta_{\rm BEC}}$}, which should be positive for the stable superfluid phase~\cite{PhysRevB.74.144517} {and should be calculated via the variational principle}.
Also, the relative phases among order parameters are determined such that $E_{\rm J}$ is minimized as $\cos(\theta_{ff}-\theta_{FF}+2\theta_{\rm BEC})=1$. 

\section{Continuity between atomic and molecular superfluids}\label{sec:3}

Let us now study the crossover behavior from the atomic $ff$ superfluid to the molecular $FF$ superfluid by increasing the closed channel molecular energy $\nu_F$.
Similar to the case of the coupled-channel system with a contact $s$-wave interaction~\cite{Ochi2022PhysRevResearch.4.013032}, the $ff$ and $FF$ pairs are coherently coupled by the pair-exchange interaction $V_{\rm SMW}$.
In this regard, $V_{\rm SMW}$ plays a crucial role for the BCS-BCS crossover between atomic and molecular superfluid states.  
Moreover, the effective coupling strength $\lambda$ is strongly enhanced at $\rho_{f}\simeq \rho_{F}$ due to the logarithmic singularity.
On the other hand, the effect of the one-body mixing $V_{\rm M}$ characterized by $\gamma$ is insignificant during the crossover between the $ff$ and $FF$ pairing states as shown in Appendix~\ref{app:gamma}.
Therefore, we take $\gamma=0$ in the following analysis, where we obtain the ground-state energy similarly to the two-band BCS model,
\begin{align}
    E_{\rm GS}
    =&\frac{1}{2}\sum_{\bm{p}}\left( \varepsilon_{\bm{p},f}+\varepsilon_{\bm{p},F} 
-E_{\bm{p},f}-E_{\bm{p},F}
\right)
+E_{\rm J}+E_{\rm Bose},
\label{eq:EGS-III}    
\end{align}
with
\begin{align}
E_{\bm{p},f}=\sqrt{\varepsilon_{\bm{p},f}^2+|\Delta_{ff}(\bm{p})|^2}, \\  E_{\bm{p},F}=\sqrt{\varepsilon_{\bm{p},F}^2+|\Delta_{FF}(\bm{p})|^2}.
\end{align} 

{
In the following subsections, we present the numerical results to demonstrate our continuity picture.
First, we derive the gap equations in Sec.~\ref{sec:IIIA}.
The relevant dimensionless parameters are then introduced in Sec.~\ref{sec:IIIB} and evaluated accordingly based on typical realistic systems or experimental data.
For specific numerical calculations, we consider $m_F/m_f=(40+23)/40=1.575$, which is relevant to the $^{23}$Na-$^{40}$K mixture~\cite{duda2021transition,duda2022longlived}.
In particular, we investigate the number densities and density-dependent pair-exchange coupling in Sec.~\ref{sec:IIIC}.
The superfluid gaps are also calculated in Sec.~\ref{sec:IIID} and a related discussion is given.
}

\subsection{Gap equations}\label{sec:IIIA}

The gap equations\footnote{In numerical analysis, it is convenient to introduce the dimensionless variables which read $m_{f,F}/m_0$, $k_{f,F,b}/k_0$, $p/k_0$, $k_{\rm cut}/k_0$,  $\Delta_{ff,FF}(\bm{p})/E_0$, $\bar{\omega}/E_0$, $\tilde{g} = g \sqrt{m_b m_f /k_0}$, $2m_{0}k_0^3\lambda$, and $2m_{0}k_0^3\lambda_{ii}$.} are obtained by taking variations of Eq.~(\ref{eq:EGS-III}) with respect to $d_{FF}$ and $d_{ff}$, respectively,
\begin{widetext}
\begin{align}
\label{eq:gapeqs}
    \left(\begin{array}{cc}
      -\lambda_{ff}-\lambda_{ff}^2\sum_{\bm{p}}\frac{p^2Y_{10}^2(\hat{\bm{p}})}{2E_{\bm{p},f}}
      -|\lambda|^2\sum_{\bm{p}}\frac{p^2Y_{10}^2(\hat{\bm{p}})}{2E_{\bm{p},F}}
      & -\lambda^*-\lambda_{ff}\lambda^*\sum_{\bm{p}}\frac{p^2Y_{10}^2(\hat{\bm{p}})}{2E_{\bm{p},f}}
      - \lambda_{FF}\lambda^*\sum_{\bm{p}}\frac{p^2Y_{10}^2(\hat{\bm{p}})}{2E_{\bm{p},F}}
      \\
      -\lambda-
      \lambda\lambda_{ff}\sum_{\bm{p}}
      \frac{p^2Y_{10}^2(\hat{\bm{p}})}{2E_{\bm{p},f}}
      -\lambda_{FF}\lambda
      \sum_{\bm{p}}\frac{p^2Y_{10}^2(\hat{\bm{p}})}{2E_{\bm{p},F}}
         & 
         -\lambda_{FF}-\lambda_{FF}^2\sum_{\bm{p}}\frac{p^2Y_{10}^2(\hat{\bm{p}})}{2E_{\bm{p},F}}
      -|\lambda|^2\sum_{\bm{p}}\frac{p^2Y_{10}^2(\hat{\bm{p}})}{2E_{\bm{p},f}}
    \end{array}\right)
    \left(
    \begin{array}{c}
         d_{ff}  \\
         d_{FF} 
    \end{array}
    \right)=\bm{0}.
\end{align}
\end{widetext}
As mentioned before, we have a momentum cutoff $k_{\rm cut}$ for the integral over $|\bm{p}|$.
 
\subsection{Dimensionless parameters}\label{sec:IIIB}

\begin{table}[b]
\caption{Summary of dimensionless parameters}
\begin{ruledtabular}
\begin{tabular}{ccc}
Parameter & Physical meaning & Numerical value \\ \hline 
$m_b/m_f$ & mass ratio & $23/40$\\
$k_{b}/k_0$ & {Eq.(\ref{eq:k_b})} & $6$\\
$k_{\rm cut}/k_0$
& cutoff momentum & $6$ \\
$ \tilde{g}= g \sqrt{m_b m_f /k_0}$ & Feshbach coupling & $O(0.01)$ \\
\end{tabular}
\end{ruledtabular}
\label{table:2}
\end{table}

Summarized in Table~\ref{table:2} are the four dimensionless parameters, their physical meanings and the actual values taken in our numerical calculations.
As a typical example, we consider the $^{23}$Na-$^{40}$K mixture ($m_b/m_f=23/40$) relevant to the recent experiments~\cite{duda2021transition,duda2022longlived}.
To estimate $k_b/k_0$, we use the scattering length $a_{bb}=g_{bb}m_b/(4\pi) = 2.75$~nm, and $\rho_b = 10^{13}-10^{15}$~cm$^{-3}$~\cite{Dalfovo1999RevModPhys.71.463}, as well as the large bosonic condensate ${\rho_b/}{\rho_0}=10-100$ similar to the Bose-polaron regime~\cite{duda2021transition}.
{
Here we employ a finite effective range $r_{bf}$ so that the system is stable with respect to the Thomas collapse. 
Meanwhile, we assume sufficiently large condensates with $g_{bb}\rho_b\gg E_0$ and $\rho_{b}\gg \rho_0$. Then many-body instabilities such as the phase separation and density collapse of the BEC due to fermion-mediated boson-boson attraction, which can occur when the fermion density exceeds a certain critical value~\cite{Marchetti2008Phys.Rev.B78.134517}, do not occur either.
}
The cutoff momentum $k_{\rm cut}$ is chosen as $k_b$ to be consistent with Eq.~(\ref{eq:k_b}).
Considering the experimental situation with several narrow Feshbach resonances~\cite{Park2012PhysRevA.85.051602}, we take $\tilde{g}$ to be of the order of $0.01$, corresponding to the largely negative effective range $k_0 r_{bf}$ of the order of  $-10^5$, which validates the present perturbative treatment with respect to $\tilde{g}$~\cite{Nishida2012PhysRevLett.109.240401,Tajima2018PhysRevA.97.043613}.

While we focus on the $^{23}$Na-$^{40}$K mixture in this paper, the present approach can also be applied to other Bose-Fermi mixtures such as $^{87}$Rb-$^{40}$K near the narrow Feshbach resonance~\cite{Goldwin2004PhysRevA.70.021601,Marco2019Science.363.853}.
Also, we note that the optical control of the effective range (associated with $g$ in this model) in the magnetic Feshbach resonance was proposed in Refs.~\cite{Wu2012PhysRevA.86.063625,Hu2020PhysRevA.101.013615}.

\subsection{Number densities and density-dependent pair-exchange coupling}\label{sec:IIIC}

\begin{figure}
    \centering
    \includegraphics[width=8.5cm]{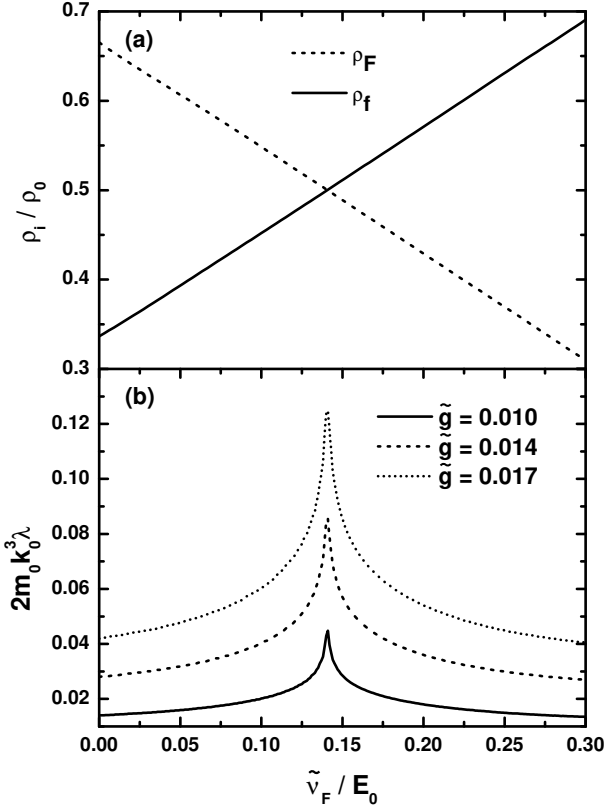}
    \caption{(a) Noninteracting number densities of atomic fermions $\rho_f/\rho_0$ and molecular fermions $\rho_F/\rho_0$ in Eq.~\eqref{nonrho}.
    (b) Dimensionless interaction strength {$2m_0 k_0^3 \lambda$} in Eq.~\eqref{eq:LAM} calculated from the non-interacting number densities.
}
    \label{fig:new5}
\end{figure}

Since we are interested in the weak-coupling BCS regime, we may use an approximation where the interaction effect on the number densities is negligible such that we can use their expressions of the non-interacting case by taking $E_f=\mu_f$ and $E_F=\mu_F$.
Specifically, the number densities of atomic and molecular fermions read
\begin{align}\label{nonrho}
    \rho_f=\frac{(2m_f\mu_f)^{3/2}}{6\pi^2}, \quad
    \rho_F=\frac{(2m_F\mu_F)^{3/2}}{6\pi^2},
\end{align}
with $\mu_F=\mu_f+\mu_b-\nu_F=\mu_f-\tilde{\nu}_F$.
Figure~\ref{fig:new5}(a) exhibits the crossover of the number fractions $\rho_f/\rho_0$ and $\rho_F/\rho_0$.
One can clearly see the decrease of $\rho_F$ and the increase of $\rho_f$ with increasing $\tilde{\nu}_F$.
While the closed-channel component is dominant for negative $\tilde{\nu}_{F}$, dissociated fermionic atoms are dominant for positive $\tilde{\nu}_{F}$.
The crossing point $\rho_f=\rho_F$ depends on the mass ratio $m_F/m_f$, but the overall structures are unchanged qualitatively.

Let us recapitulate the density-dependent pair-exchange coupling $\lambda$ to be used in the present numerical calculations, where
\begin{align}
\label{eq:LAM}
\lambda= -\frac{3\pi {m_bg^2}(k_f^2+k_F^2)}{4k_f^3k_F^3}
    \ln\left[\frac{\frac{\alpha^2\bar{\omega}^2}{v_b^2}+(k_f-k_F)^4}{\frac{\alpha^2\bar{\omega}^2}{v_b^2}+(k_f+k_F)^4}\right].
\end{align}
Here, we take $\theta_{ff}-\theta_{FF}+2\theta_{\rm BEC}=0$. 
{Note that $\lambda$ is always positive.}
To see the region where the pair-exchange process is significant, we plot {$2m_0k_0^3\lambda$} as a function of the reduced molecular energy $\tilde{\nu}_F$  in Fig.~\ref{fig:new5}(b), where $k_f$ and $k_F$ are evaluated by using {Eqs.~\eqref{eq17} and} \eqref{nonrho}.
There $\lambda$ shows a sharp peak at $\rho_f=\rho_F$ due to the logarithmic singularity.
In particular, the maximum value at $\tilde{\nu}_F/E_0\simeq 0.13$ reads
\begin{align}
    \lambda_{\rm max}\simeq
    \frac{3\pi m_b g^2}{k_f^4}\ln\left(\frac{16\pi v_b}{m_fm_F\zeta g^2} \right) \quad (k_f=k_F).
\end{align}
Because the BCS superfluid is supported by the Fermi energy of each component (i.e., $E_f$ and $E_F$) and the pair-exchange coupling $\lambda$,
the dominant pairing state is expected to gradually change from the atomic BCS state to the molecular BCS state around $\tilde{\nu}_F/E_0\simeq 0.13$ without any phase transition.
In the following, we focus on this crossover regime ($0.06\lesssim \tilde{\nu}_F/E_0\lesssim 0.24$).

\subsection{Superfluid gaps}\label{sec:IIID}

\begin{figure}
    \centering
    \includegraphics[width=8.5cm]{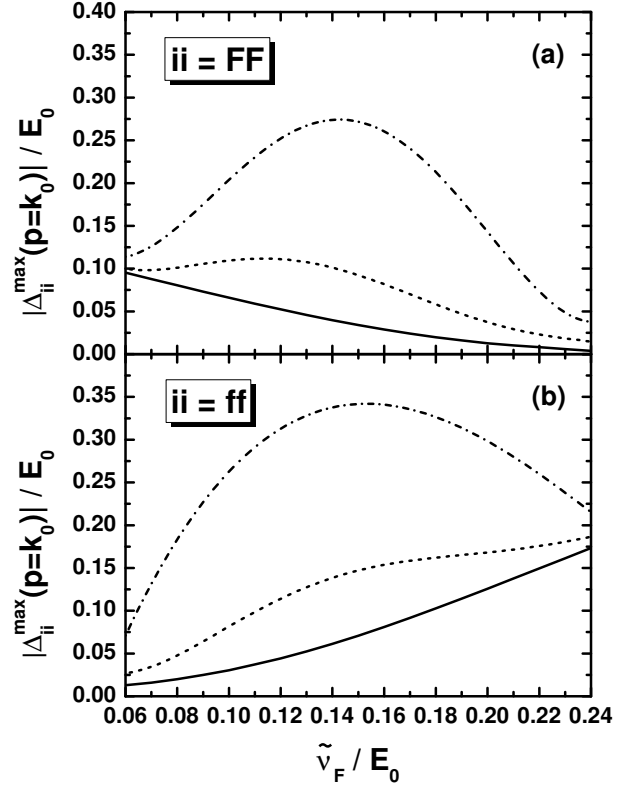}
    \caption{Pairing gaps $\Delta_{FF}^{\rm max}(|\bm{p}|=k_0)/E_{0}$ and $\Delta_{ff}^{\rm max}(|\bm{p}|=k_0)/E_{0}$ with and without $V_{Fbf}$ shown as a function of $\tilde{\nu}_F$, where $1/(v_jk_0^3)$ is taken as $-0.3$ (with $2m_0k_0^3\lambda_{ff}=-3.89$ and $2m_0k_0^3\lambda_{FF}=-2.47$).
    The case without $V_{Fbf}$ ($\tilde{g}=0$) is shown by solid lines, while the cases $\tilde{g}=0.010$ (with $2m_0k_0^3\lambda\sim[0.017,0.045]$) and $\tilde{g}=0.017$ (with $2m_0k_0^3\lambda\sim[0.050,0.125]$) are shown by dashed and dash-dotted lines, respectively.
}
    \label{fig:4}
\end{figure}
From Eqs.~\eqref{eq:delta_ff} and \eqref{eq:delta_FF} it is found that the pairing gap exhibits a maximum in a direction $\hat{\bm{p}}$ with $Y_{10}(\bm{\hat{p}})=\sqrt{\frac{3}{4\pi}}$.
In this regard, we define 
$\Delta_{FF}^{\rm max}(|\bm{p}|=k_0)=\sqrt{\frac{3}{4\pi}}k_0\left[|\lambda| |d_{ff}|+|\lambda_{FF}||d_{FF}|\right]$ and $\Delta_{ff}^{\rm max}(|\bm{p}|=k_0)=\sqrt{\frac{3}{4\pi}}k_0\left[|\lambda| |d_{FF}|+|\lambda_{ff}||d_{ff}|\right]$ to examine the pairing effect.
Dimensionless pairing gaps $\Delta_{FF}^{\rm max}(|\bm{p}|=k_0)/E_{0}$ and $\Delta_{ff}^{\rm max}(|\bm{p}|=k_0)/E_{0}$ with and without $V_{\rm SMW}$ are plotted in Fig.~\ref{fig:4}, where the intracomponent couplings are taken as $2m_0k_0^3\lambda_{ff}=-3.89$ and $2m_0k_0^3\lambda_{FF}=-2.47$.
In the absence of $V_{\rm SMW}$, the pairing gaps are decoupled from each other and their magnitudes are purely determined by $\lambda_{ii}$
$(i=f,F)$ as
\begin{align}
1+\lambda_{ii}\sum_{\bm{p}}\frac{p^2Y_{10}^2(\hat{\bm{p}})}{2E_{\bm{p},i}}=0.
\end{align}
While $\Delta_{ff}^{\rm max}(|\bm{p}|=k_0)$ increases with increasing $\tilde{\nu}_F$, $\Delta_{FF}^{\rm max}(|\bm{p}|=k_0)$ decreases, reflecting the behavior of $\rho_f$ and $\rho_F$.
{
In order to achieve the weak-coupling regime in numerical calculations, we let the dimensionless Feshbach coupling $\tilde{g}$ be of the order of $0.01$ leading to a superfluid gap sufficiently smaller than the Fermi energy scale $E_0$.
}
In the presence of $V_{\rm SMW}$ where $\tilde{g}\sim O(0.01)$ is taken, 
we calculate the pairing gaps by using the perturbative treatment with respect to $2m_0k_0^3\lambda$, which is sufficiently small compared to $2m_0k_0^3\lambda_{ff}=-3.89$ and $2m_0k_0^3\lambda_{FF}=-2.47$ as shown in Fig.~\ref{fig:new5}(b).
It is found that both pairing gaps $\Delta_{FF}^{\rm max}(|\bm{p}|=k_0)/E_{0}$ and $\Delta_{ff}^{\rm max}(|\bm{p}|=k_0)/E_{0}$ increase compared to the case with $\tilde{g}=0$. 
The enhancement of $\Delta_{FF(ff)}(\bm{p})$ is associated with the increase of the pair-exchange interaction $\lambda$ due to the logarithmic singularity at $k_{f}=k_{F}$ (i.e., $\rho_f=\rho_F$).
This indicates that the long-range property of $U_{\rm SMW}$ associated with the low-energy Bogoliubov phonons strongly assists these fermionic superfluid states.
At the same time, the Landau damping suppresses the logarithmic divergence due to the mass-imbalance between atomic and molecular fermions, and hence we obtain the finite values of $\Delta_{FF}(\bm{p})$ and $\Delta_{ff}(\bm{p})$ in the entire crossover region.

Apart from the long-range nature of the interaction, this result is also consistent with the fact that the pairing gaps are largely enhanced by the resonant pair scattering~\cite{Ochi2022PhysRevResearch.4.013032,PhysRevB.82.184528,mazziotti2021resonant}.
Because the Landau damping factor is proportional to $g^2$ at the leading order, the associated $\alpha$ can also be small and hence the enhancement of both pairing gaps can be anticipated even for small $g$, as shown in Fig.~\ref{fig:4}.
Although our weak-coupling approximation is valid where the pairing gaps are sufficiently small, one can see that both pairing gaps are strongly enhanced when the value of $g$ increases, as expected regardless of the weak-coupling approximation.
In this sense, the pair-exchange process would be much more significant for larger Feshbach couplings.
Note that perturbative calculations should be understood as qualitative near the resonance region where there is a large enhancement of the gaps.  Further theoretical development is needed to perform quantitative calculations, which is left for future study.

Regarding the stability of the system, the Josephson coupling energy $E_{\rm J}$ should be positive throughout the crossover.
This fact can be understood from Eq.~\eqref{eq:EGS-III}, where the first term $\frac{1}{2}\sum_{\bm{p}}[\varepsilon_{\bm{p},f}+\varepsilon_{\bm{p},F}-E_{\bm{p},f}-E_{\bm{p},F}]$ is a decreasing function with respect to $|d_{ff(FF)}|$ at $T=0$.
To find the global minimum, $E_{\rm J}$ should be an increasing function of $|d_{ff(FF)}|$ (otherwise, the system would collapse to an infinitely large negative energy state with $|d_{ff(FF)}|\rightarrow\infty$).
We confirm that $E_{\rm J}>0$ is satisfied during the whole investigating region. 
We note that at finite temperature the thermal energy is also involved in the quadratic term with respect to the order parameters of the Ginzburg-Landau energy~\cite{PhysRevB.74.144517}.
In this regard, the stability condition ($E_{\rm J}>0$) will be modified at finite temperature.

\section{Discussion}\label{sec:4}

The experimental observation of the continuity between two superfluid states is feasible by measuring the excitation gaps of each component through the radio-frequency spectroscopy~\cite{Schirotzek2008PhysRevLett.101.140403} as well as the momentum-resolved photoemission spectroscopy~\cite{stewart2008Nature.454.744,Sagi2015PhysRevLett.114.075301}.
Measuring the enhancements of two gaps with changing $\tilde{\nu}_F$ near the Feshbach resonance can provide evidence of the coupling between two Fermi atomic and molecular superfluid states associated with pair-exchange interaction involving the long-range nature with logarithmic singularity.
Such an enhancement becomes more significant when the Feshbach coupling is larger.

To see a more direct consequence of the pair-exchange coupling, it is helpful to borrow from the physics of multiband superconductors.
In particular, one may find the intrinsic Josephson effect~\cite{Iskin2016PhysRevA.94.011604} in the presence of the pair-exchange coupling.
The time evolution of the average number-density difference is given by the Heisenberg equation
\begin{align}
\label{eq:Josephson}
    &\frac{d (\rho_f-\rho_F)}{dt}=i\langle[H_{\rm eff}^{\rm MF},\hat{N}_f-\hat{N}_F] \rangle\cr
    &\quad \simeq-4|\lambda||d_{FF}||d_{ff}|\sin(\theta_{ff}-\theta_{FF}+2\theta_{\rm BEC}),
\end{align}
where $\hat{N}_{f}=\sum_{\bm{p}}f_{\bm{p}}^\dag f_{\bm{p}}$ and $\hat{N}_{F}=\sum_{\bm{p}}F_{\bm{p}}^\dag F_{\bm{p}}$ are the number operators.
In Eq.~(\ref{eq:Josephson}) the statistical average $\langle \cdots \rangle$ is approximately evaluated within the thermal equilibrium.
In this way, the Josephson current occurs when the relative phase is modulated from the equilibrium value $\theta_{ff}-\theta_{FF}+2\theta_{\rm BEC}=0$.
This relative-phase-dependent term is proportional to $|\lambda|$
and therefore such a behavior of the number density difference between atoms and molecules can be evidence of the pair-exchange coupling and the resulting crossover.
We note that the Josephson effect of a two-component superfluid Fermi gas has already been observed experimentally~\cite{valtolina2015josephson}.

Another interesting feature of the multiband superconductors is the emergence of the Leggett mode~\cite{Leggett1966PTP.36.901}, which is the out-of-phase collective excitation of two order parameters.
Like the massless Leggett mode discussed in three-band superconductors~\cite{Lin2012PhysRevLett.108.177005}, a similar collective mode may appear in the present system due to the three coupled phases
{$\theta_{ff}$, $\theta_{FF}$, and $\theta_{\rm BEC}$.}
Because the collective modes of Fermi superfluids were experimentally measured in cold-atom systems (e.g.,the Nambu-Goldstone mode~\cite{Hoinka2017Nat.Phys.13.943} and {the Higgs mode}~\cite{Behrle2018Nat.Phys.14.781}),
the analog of {the Leggett mode} associated with atom-atom and molecule-molecule pairing order parameters in a Bose-Fermi mixture should be accessible in the future.
In addition, in the case with heavy bosons (and hence heavy molecular fermions), the high-temperature superconducting mechanism discussed in the context of the FeSe multiband superconductor may appear due to the realization of the shallow band with screened pairing fluctuations~\cite{Tajima2019PhysRevB.99.180503,Salasnich2019PhysRevB.100.064510}.
These nontrivial effects which are specific for multiband fermionic superfluid can also be utilized to confirm the continuity picture for atomic and molecular Fermi superfluids in a Bose-Fermi mixture.

Here we mention the case where the bosonic density $\rho_b$ is comparable to $\rho_f$ and hence the atom-dimer coupling effect on $\rho_b$ is not negligible.
As {a first} approximation, we can incorporate these effects by considering the mean-field framework of bosons as 
${\partial E_{\rm GS}}/{\partial \rho_b}=0$ where $\rho_b$ is self-consistently determined.
It can be rewritten as
\begin{align}
\label{eq:withBEC}
    -\mu_b+g_{bb}\rho_b+\frac{\partial \lambda}{\partial \rho_b}\frac{\partial E_{\rm Fermi}}{\partial \lambda}
    +\frac{g}{2 \sqrt{\rho_b}}\frac{\partial E_{\rm Fermi}}{\partial \gamma}=0,
\end{align}
{where the fermionic energy density $E_{\rm Fermi}$ is defined through}  $E_{\rm GS}=E_{\rm Bose}+E_{\rm Fermi}$.
In the case with large condensates $\rho_b\gg \rho_{f(F)}$, we may recover the situation that we demonstrated above, because the last two terms in Eq.~(\ref{eq:withBEC}) vanish.
Exploring physical properties around the tricritical point in the system with $\rho_f=\rho_b$ observed in Ref.~\cite{duda2021transition} is left for interesting future work.
Nevertheless, the present continuity picture may be qualitatively valid even in the case with $\rho_f\simeq\rho_b$ once the coexistence of $FF$ and $ff$ pairing states and the bosonic condensates is found around $\nu_F=0$.
Moreover, the present theoretical model can be easily extended to a spin-$\frac{1}{2}$ Bose-Fermi mixture with the $s$-wave pairing, which is more relevant to dense QCD~\cite{Maeda2009PhysRevLett.103.085301}.

\section{Summary}\label{sec:5}

We have investigated theoretically the continuity between atomic and molecular Fermi superfluids in a Bose-Fermi mixture near the heteronuclear Feshbach resonance.
Considering the perturbative regime with respect to the Feshbach atom-dimer coupling, we have developed {a multicomponent} superfluid theory. 
The effective mean-field Hamiltonian for the atom-atom and molecule-molecule pairing states is equivalent to the two-band BCS Hamiltonian originally proposed by Suhl, Matthius, and Walker for superconductors with overlapping bands, except for the pairing symmetry.

In particular, we found that the pair-exchange term for the atom-atom and molecule-molecule pairing states occurs through the anomalous propagator of the bosonic component.
Such a pair-exchange coupling can be dramatically enhanced in the crossover regime due to the long-range property of the mediated interaction.
We have demonstrated numerically the continuity picture within the simplified models.
The existence of the pair-exchange coupling can be probed via the intrinsic Josephson effect, where the difference between the atomic and molecular number densities is associated with the relative phases, $\theta_{ff}-\theta_{FF}+2\theta_{\rm BEC}$.
The present results towards the quantum simulation of dense QCD using atomic Bose-Fermi mixture imply that the continuity between the dibaryon and diquark condensates can also be understood in terms of the two-band-like superfluid theory.

For a quantitative comparison with experiments, further investigations with realistic interactions and wider ranges of density fractions are needed.
Moreover, it is also important to consider the strong-coupling effects associated with higher-order terms of the Feshbach coupling.
In such a case, we need to consider other multibody cluster states such as mass-imbalanced atom-molecule pairing beyond the BCS framework.
It would also be interesting to address the speed of sound in the crossover regime~\cite{Shi2022}.

\begin{acknowledgments}
The authors are grateful to Xingyan Chen and Kohei Kato for fruitful discussions about the experiments.
Y.G. was supported by RIKEN Junior Research Associate Program.
H.T. acknowledges the JSPS Grants-in-Aid for Scientific Research under Grant No.~18H05406.
T.H. acknowledges the JSPS Grants-in-Aid for Scientific Research under Grant No.~18H05236
H.L. acknowledges the JSPS Grant-in-Aid for Early-Career Scientists under Grant No.~18K13549, the JSPS Grant-in-Aid for Scientific Research (S) under Grant No.~20H05648, and the RIKEN Pioneering Project: Evolution of Matter in the Universe.
\end{acknowledgments}

\appendix

{
\section{Power counting in the numerator of Eqs.~\eqref{eq:U-SMW} and \eqref{eq:U-PhM}}\label{app:pc}}

{
Here, we perform the power counting with respect to the fermionic momenta $k$ and $k'$ for the term $2i\omega\Gamma(\bm{q},\omega)$ in the numerator of Eq.~\eqref{eq:U-SMW} in order to obtain Eq.~\eqref{eq:approx} with the large bosonic momentum scale $k_{i=f,F}/k_b\ll 1$.
In detail, dividing the numerator of Eq.~\eqref{eq:U-SMW} by the squared bosonic energy scale $(g_{bb}\rho_b)^2$, we have
\begin{align}
    &\frac{\bar{\omega}^2-E_{\bm{q},b}^2+\Gamma^2(\bar{q},\bar{\omega})-2i\bar{\omega}\Gamma(\bar{q},\bar{\omega})}{(g_{bb}\rho_b)^2}\nonumber\\
    =\,&\frac{-v_{b}^2\bar{q}^2+\Gamma^2(\bar{q},\bar{\omega})}{(g_{bb}\rho_b)^2}+O(k_{i}k_{j}k_{\ell}/k_b^3) \quad (i,j,\ell=f,F),
\end{align}
where we used the relations
\begin{align}
    \frac{\bar{\omega}^2}{(g_{bb}\rho_b)^2}
&=\left(\frac{m_b}{m_f}\frac{k^2}{k_b^2}-\frac{m_b}{m_F}\frac{k'^2}{k_b^2}\right)^2\cr
    &\equiv O(k_{i}k_{j} k_{\ell}/k_b^3),
\end{align}
\begin{align}
    \frac{E_{\bar{q},b}^2}{(g_{bb}\rho_b)^2}&=\frac{v_{b}^2\bar{q}^2}{(g_{bb}\rho_b)^2}+O(\bar{q}^3/k_b^3)\cr
    &\equiv \frac{v_{b}^2\bar{q}^2}{(g_{bb}\rho_b)^2}
    +O(k_{i}k_{j}k_{\ell}/k_b^3),
\end{align}
\begin{align}
    \frac{\Gamma^2(\bar{q},\bar{\omega})}{(g_{bb}\rho_b)^2}
    =\alpha^2 \frac{|\bar{\omega}|^2}{\bar{q}^2(g_{bb}\rho_b)^2}\equiv O(k_ik_j/k_b^2),
\end{align}
and
\begin{align}
    \frac{2i\bar{\omega}\Gamma(\bar{q},\bar{\omega})}{(g_{bb}\rho_b)^2}=2i
    \frac{\bar{\omega}}{g_{bb}\rho_b}
        \frac{\Gamma(\bar{q},\bar{\omega})}{g_{bb}\rho_b}
        \equiv O(k_{i} k_{j} k_{\ell}/k_b^3).
\end{align}
Furthermore, as for the scattering process $fF \rightarrow Ff$, the numerator of Eq.~\eqref{eq:U-PhM} with the same power counting reads
\begin{align}
&\frac{[\omega-E_{\bm{q},b}-i\Gamma(\bm{q},\omega)][\omega+E_{\bm{q},b}-i\Gamma(\bm{q},\omega)]}{(g_{bb}\rho_b)^2}\nonumber\\
=\,&\frac{v_b^2q^2+\Gamma^2(q,\omega)}{(g_{bb}\rho_b)^2} +O(k_i k_j k_\ell/k_b^3),  
\end{align}
where we used
\begin{align}
\frac{\omega+\varepsilon_{\bm{q}}+g_{bb}\rho_b+i\Gamma(\bm{q},\omega)}{g_{bb}\rho_b}=1+O(k_i/k_b).
\end{align}
The resulting effective interaction 
\begin{align}
U_{\rm PM}(\bm{q},\omega)=-g_{bb}\rho_b\frac{1}{v_b^2q^2+\alpha^2\frac{\omega^2}{q^2}}
\end{align}
does not show the infrared singularity at $q \rightarrow 0$ due to the existence of $\alpha$ (i.e., the Landau damping). 
}

\section{Derivation of Eq.~(\ref{eq:U1})}\label{App:U1}

In this appendix, we show the derivation of Eq.~(\ref{eq:U1}).
The projection of $U_{\rm SMW}(\bar{q},\bar{\omega})$ to the $p$-wave ($\ell=1$) component is given by
\begin{align}
 U_{1 m}(k_f,k_F)&= \frac{1}{4\pi}\int d\Omega_{k}\int d\Omega_{k'}
    U_{\rm SMW}(\bar{q},\bar{\omega})\cr
    &\ \ \ \times Y_{1m}^*(\hat{\bm{k}})Y_{1m}(\hat{\bm{k}}'),
\end{align}
where $\Omega_{k}$ is the solid angle with respect to $\bm{k}$.
One can see that the $m=\pm 1$ components disappear after the angular integration.
Note that $\bar{q}$ is a function of $\theta_{\bm{k}\bm{k}'}\equiv\hat{\bm{k}}\cdot\hat{\bm{k}}'$.
By taking the axis along the $\bm{k}$ direction, we can reduce the angular integration as
\begin{align}
\label{eq:a2}
        U_{1 0}(k_f,k_F)
    &=\frac{3}{2}\int_{0}^{\pi} d\theta_{\bm{k}\bm{k}'}\sin\theta_{\bm{k}\bm{k}'}\cos\theta_{\bm{k}\bm{k}'} 
    U_{\rm SMW}(\bar{q},\bar{\omega}).
\end{align}
The $\theta_{\bm{k}\bm{k}'}$ integration in Eq.~(\ref{eq:a2}) can be performed analytically as
\begin{widetext}
\begin{align}
\label{eq:A3}
    U_{1 0}(k_f,k_F)
    \simeq&-\frac{3m_bg^2e^{2i\theta_{\rm BEC}}}{2k_fk_F}
    -\frac{3m_bg^2e^{2i\theta_{\rm BEC}}(k_f^2+k_F^2)}{16k_f^2k_F^2}
    \ln\left[\frac{\alpha^2\bar{\omega}^2+v_b^2(k_f-k_F)^4}{\alpha^2\bar{\omega}^2+v_b^2(k_f+k_F)^4}
    \right]
    \cr
    &-\frac{3m_bg^2e^{2i\theta_{\rm BEC}}\alpha^2\bar{\omega}^2}{4k_fk_F}
    \left[
    \frac{1 }{\alpha^2\bar{\omega}^2+v_b^2(k_f+k_F)^4}
    +\frac{1 }{\alpha^2\bar{\omega}^2+v_b^2(k_f-k_F)^4}
    \right]
    \cr
    &+\frac{3m_bg^2e^{2i\theta_{\rm BEC}}\alpha\bar{\omega}}{4v_bk_f^2k_F^2}\left[
    \arctan\left(\frac{v_b(k_f+k_F)^2}{\alpha\bar{\omega}}\right)
    -\arctan\left(\frac{v_b(k_f-k_F)^2}{\alpha\bar{\omega}}\right)
    \right],
\end{align}
\end{widetext}
where we used Eq.~\eqref{eq:approx} in Eq.~\eqref{eq:a2}.
In the weak-coupling limit (i.e., $g\rightarrow 0$) near $k_f=k_F$, the logarithmic term become extremely large so that for simplicity we keep only this singular term as shown in Eq.~\eqref{eq:U1}.

\section{One-body mixing effect}\label{app:gamma}

To see the more relevant situation for an ultracold atomic Bose-Fermi mixture near the Feshbach resonance,
one may consider the nonzero interband impurity effect (i.e., one-body mixing) occurring at the lowest order of $g$.  
The gap equations {in this case} can be obtained from
\begin{widetext}
\begin{subequations}
\begin{align}
    \frac{\partial E_{\rm GS}}{\partial d_{ff}^*}
    =\,&-\frac{1}{2}({\lambda^*d_{FF} +\lambda_{ff}d_{ff}})\nonumber\\
   & -\frac{1}{2}\sum_{\bm{p}}
    \frac{1}{4e_1^{+}}\left\{\left[
1   -\frac{(\mathcal{A}+2\gamma^2)-2\varepsilon_{\bm{p},F}^2-2\left|\Delta_{FF}(\bm{p})\right|^2}
   {\sqrt{(\mathcal{A}+2\gamma^2)^2-4\mathcal{B}}}
    \right]\Delta_{ff}(\bm{p})
    +\frac{2\gamma^2\Delta_{FF}(\bm{p})}
   {\sqrt{(\mathcal{A}+2\gamma^2)^2-4\mathcal{B}}}\right\} \frac{\partial \Delta_{ff}^*(\bm{p})}{\partial d_{ff}^*}\nonumber\\
   & -\frac{1}{2}\sum_{\bm{p}}
    \frac{1}{4e_2^{+}}\left\{\left[
1   +\frac{(\mathcal{A}+2\gamma^2)-2\varepsilon_{\bm{p},F}^2-2\left|\Delta_{FF}(\bm{p})\right|^2}
   {\sqrt{(\mathcal{A}+2\gamma^2)^2-4\mathcal{B}}}
    \right]\Delta_{ff}(\bm{p})
    -\frac{2\gamma^2\Delta_{FF}(\bm{p})}
   {\sqrt{(\mathcal{A}+2\gamma^2)^2-4\mathcal{B}}}\right\} \frac{\partial \Delta_{ff}^*(\bm{p})}{\partial d_{ff}^*}
   \nonumber\\
   &-\frac{1}{2}\sum_{\bm{p}}
    \frac{1}{4e_1^{+}}\left\{\left[
1   -\frac{(\mathcal{A}+2\gamma^2)-2\varepsilon_{\bm{p},f}^2-2\left|\Delta_{ff}(\bm{p})\right|^2}
   {\sqrt{(\mathcal{A}+2\gamma^2)^2-4\mathcal{B}}}
    \right]\Delta_{FF}(\bm{p})
    +\frac{2\gamma^2\Delta_{ff}(\bm{p})}
   {\sqrt{(\mathcal{A}+2\gamma^2)^2-4\mathcal{B}}}\right\}
   \frac{\partial \Delta_{FF}^*(\bm{p})}{\partial d_{ff}^*}
   \nonumber\\
   & -\frac{1}{2}\sum_{\bm{p}}
    \frac{1}{4e_2^{+}}\left\{\left[
1   +\frac{(\mathcal{A}+2\gamma^2)-2\varepsilon_{\bm{p},f}^2-2\left|\Delta_{ff}(\bm{p})\right|^2}
   {\sqrt{(\mathcal{A}+2\gamma^2)^2-4\mathcal{B}}}
    \right]\Delta_{FF}(\bm{p})
    -\frac{2\gamma^2\Delta_{ff}(\bm{p})}
   {\sqrt{(\mathcal{A}+2\gamma^2)^2-4\mathcal{B}}}\right\}
   \frac{\partial \Delta_{FF}^*(\bm{p})}{\partial d_{ff}^*}\nonumber\\
   =\,&0, \\
    \frac{\partial E_{\rm GS}}{\partial d_{FF}^*}
    =\,&-\frac{1}{2}({\lambda d_{ff}+\lambda_{FF}d_{FF}})\nonumber\\
    & -\frac{1}{2}\sum_{\bm{p}}
    \frac{1}{4e_1^{+}}\left\{\left[
1   -\frac{(\mathcal{A}+2\gamma^2)-2\varepsilon_{\bm{p},F}^2-2\left|\Delta_{FF}(\bm{p})\right|^2}
   {\sqrt{(\mathcal{A}+2\gamma^2)^2-4\mathcal{B}}}
    \right]\Delta_{ff}(\bm{p})
    +\frac{2\gamma^2\Delta_{FF}(\bm{p})}
   {\sqrt{(\mathcal{A}+2\gamma^2)^2-4\mathcal{B}}}\right\} \frac{\partial \Delta_{ff}^*(\bm{p})}{\partial d_{FF}^*}\nonumber\\
   & -\frac{1}{2}\sum_{\bm{p}}
    \frac{1}{4e_2^{+}}\left\{\left[
1   +\frac{(\mathcal{A}+2\gamma^2)-2\varepsilon_{\bm{p},F}^2-2\left|\Delta_{FF}(\bm{p})\right|^2}
   {\sqrt{(\mathcal{A}+2\gamma^2)^2-4\mathcal{B}}}
    \right]\Delta_{ff}(\bm{p})
    -\frac{2\gamma^2\Delta_{FF}(\bm{p})}
   {\sqrt{(\mathcal{A}+2\gamma^2)^2-4\mathcal{B}}}\right\} \frac{\partial \Delta_{ff}^*(\bm{p})}{\partial d_{FF}^*}
   \nonumber\\
    &-\frac{1}{2}\sum_{\bm{p}}
    \frac{1}{4e_1^{+}}\left\{\left[
1   -\frac{(\mathcal{A}+2\gamma^2)-2\varepsilon_{\bm{p},f}^2-2\left|\Delta_{ff}(\bm{p})\right|^2}
   {\sqrt{(\mathcal{A}+2\gamma^2)^2-4\mathcal{B}}}
    \right]\Delta_{FF}(\bm{p})
    +\frac{2\gamma^2\Delta_{ff}(\bm{p})}
   {\sqrt{(\mathcal{A}+2\gamma^2)^2-4\mathcal{B}}}\right\}
   \frac{\partial \Delta_{FF}^*(\bm{p})}{\partial d_{FF}^*}
   \nonumber\\
   & -\frac{1}{2}\sum_{\bm{p}}
    \frac{1}{4e_2^{+}}\left\{\left[
1   +\frac{(\mathcal{A}+2\gamma^2)-2\varepsilon_{\bm{p},f}^2-2\left|\Delta_{ff}(\bm{p})\right|^2}
   {\sqrt{(\mathcal{A}+2\gamma^2)^2-4\mathcal{B}}}
    \right]\Delta_{FF}(\bm{p})
    -\frac{2\gamma^2\Delta_{ff}(\bm{p})}
   {\sqrt{(\mathcal{A}+2\gamma^2)^2-4\mathcal{B}}}\right\}
   \frac{\partial \Delta_{FF}^*(\bm{p})}{\partial d_{FF}^*}\nonumber\\
   =\,&0,
\end{align}
\end{subequations}
where
\begin{align}
\frac{\partial \Delta_{ff}^*(\bm{p})}{\partial d_{ff}^*}=
pY_{10}(\hat{\bm{p}})\lambda_{ff}^*
,\quad
\frac{\partial \Delta_{FF}^*(\bm{p})}{\partial d_{ff}^*}=
pY_{10}(\hat{\bm{p}})\lambda^*
,\quad
\frac{\partial \Delta_{ff}^*(\bm{p})}{\partial d_{FF}^*}=
pY_{10}(\hat{\bm{p}})\lambda
,\quad
\frac{\partial \Delta_{FF}^*(\bm{p})}{\partial d_{FF}^*}=
pY_{10}(\hat{\bm{p}})\lambda_{FF}^*
.
\end{align}
\end{widetext}
The one-body mixing strength is related to the dimensionless values given by $m_0^2g^2/k_0$ and $\rho_b/\rho_0$ as
\begin{align}
    \frac{\gamma}{E_0}
    =\frac{2}{\pi\sqrt{6}}\left(\frac{m_0^2  \tilde{g}^2 }{m_b m_f}\right)^{1/2}
      \sqrt{\frac{\rho_b}{\rho_0}}.
\end{align}
For example, for the case of $\tilde{g} =0.010$, since we consider the case with the large bosonic condensate as $\frac{\rho_b}{\rho_0}=10-100$ (e.g., Bose-polaron regime in Ref.~\cite{duda2021transition}),
the resulting $\gamma/E_{0}$ reads
\begin{align}\label{gammacc}
    0.007 \lesssim\frac{\gamma}{E_0}\lesssim 0.021. 
\end{align}

\begin{figure}
    \centering
    \includegraphics[width=8.5cm]{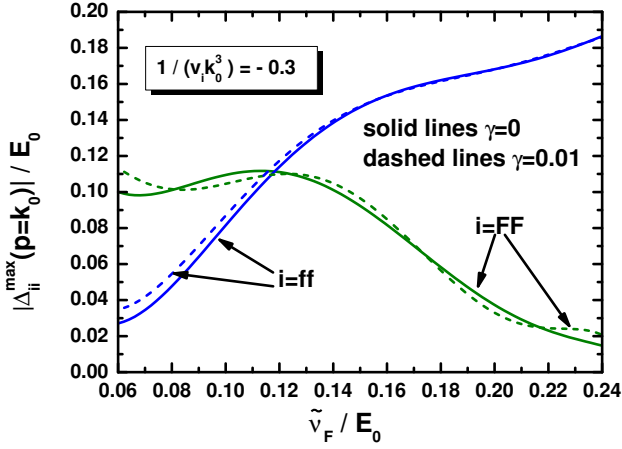}
    \caption{Pairing gaps $\Delta_{FF}^{\rm max}(|\bm{p}|=k_0)/E_{0}$ and $\Delta_{ff}^{\rm max}(|\bm{p}|=k_0)/E_{0}$ as a function of $\tilde{\nu}_F/E_0$ with $\gamma=0$ and $0.10$, where $1/(v_ik_0^3)$ is taken as $-0.3$.
    Here $\tilde{g}$ is taken as $0.01$.
}
    \label{fig:7}
\end{figure}
Figure~\ref{fig:7} shows the superfluid order parameters $|\Delta_{FF}^{\rm max}(p=k_0)|$ and $|\Delta_{ff}^{\rm max}(p=k_0)|$ near the Feshbach resonance with $\gamma/E_0=0$ and $0.01$, respectively.
The solutions are obtained by using a method similar to that shown in Fig.~\ref{fig:4}.
Since sufficiently small $g$ (weak coupling) is adopted in this work, which also restricts the corresponding $\gamma$ to be small [Eq.~\eqref{gammacc}], the difference among the results with different $\gamma$ is not qualitatively significant.
In addition, while nonzero $\gamma$ partially modifies the magnitude of the pairing gaps, their qualitative behaviors are similar to the results with $\gamma=0$. Thus we found that the crossover picture for atomic and molecular Fermi superfluids is unchanged qualitatively by the one-body mixing effect. 


\begin{thebibliography}{61}%
\makeatletter
\providecommand \@ifxundefined [1]{%
 \@ifx{#1\undefined}
}%
\providecommand \@ifnum [1]{%
 \ifnum #1\expandafter \@firstoftwo
 \else \expandafter \@secondoftwo
 \fi
}%
\providecommand \@ifx [1]{%
 \ifx #1\expandafter \@firstoftwo
 \else \expandafter \@secondoftwo
 \fi
}%
\providecommand \natexlab [1]{#1}%
\providecommand \enquote  [1]{``#1''}%
\providecommand \bibnamefont  [1]{#1}%
\providecommand \bibfnamefont [1]{#1}%
\providecommand \citenamefont [1]{#1}%
\providecommand \href@noop [0]{\@secondoftwo}%
\providecommand \href [0]{\begingroup \@sanitize@url \@href}%
\providecommand \@href[1]{\@@startlink{#1}\@@href}%
\providecommand \@@href[1]{\endgroup#1\@@endlink}%
\providecommand \@sanitize@url [0]{\catcode `\\12\catcode `\$12\catcode
  `\&12\catcode `\#12\catcode `\^12\catcode `\_12\catcode `\%12\relax}%
\providecommand \@@startlink[1]{}%
\providecommand \@@endlink[0]{}%
\providecommand \url  [0]{\begingroup\@sanitize@url \@url }%
\providecommand \@url [1]{\endgroup\@href {#1}{\urlprefix }}%
\providecommand \urlprefix  [0]{URL }%
\providecommand \Eprint [0]{\href }%
\providecommand \doibase [0]{https://doi.org/}%
\providecommand \selectlanguage [0]{\@gobble}%
\providecommand \bibinfo  [0]{\@secondoftwo}%
\providecommand \bibfield  [0]{\@secondoftwo}%
\providecommand \translation [1]{[#1]}%
\providecommand \BibitemOpen [0]{}%
\providecommand \bibitemStop [0]{}%
\providecommand \bibitemNoStop [0]{.\EOS\space}%
\providecommand \EOS [0]{\spacefactor3000\relax}%
\providecommand \BibitemShut  [1]{\csname bibitem#1\endcsname}%
\let\auto@bib@innerbib\@empty
\bibitem [{\citenamefont {Ketterle}\ and\ \citenamefont
  {Zwierlein}(2008)}]{ketterle2008NuovoCimento}%
  \BibitemOpen
  \bibfield  {author} {\bibinfo {author} {\bibfnamefont {W.}~\bibnamefont
  {Ketterle}}\ and\ \bibinfo {author} {\bibfnamefont {M.~W.}\ \bibnamefont
  {Zwierlein}},\ }\bibfield  {title} {\bibinfo {title} {Making, probing and
  understanding ultracold fermi gases},\ }\href
  {https://doi.org/10.1393/ncr/i2008-10033-1} {\bibfield  {journal} {\bibinfo
  {journal} {Riv. Nuovo Cim.}\ }\textbf {\bibinfo {volume} {31}},\ \bibinfo
  {pages} {247} (\bibinfo {year} {2008})}\BibitemShut {NoStop}%
\bibitem [{\citenamefont {Hadzibabic}\ \emph {et~al.}(2002)\citenamefont
  {Hadzibabic}, \citenamefont {Stan}, \citenamefont {Dieckmann}, \citenamefont
  {Gupta}, \citenamefont {Zwierlein}, \citenamefont {G\"orlitz},\ and\
  \citenamefont {Ketterle}}]{Hadzibabic2002PhysRevLett.88.160401}%
  \BibitemOpen
  \bibfield  {author} {\bibinfo {author} {\bibfnamefont {Z.}~\bibnamefont
  {Hadzibabic}}, \bibinfo {author} {\bibfnamefont {C.~A.}\ \bibnamefont
  {Stan}}, \bibinfo {author} {\bibfnamefont {K.}~\bibnamefont {Dieckmann}},
  \bibinfo {author} {\bibfnamefont {S.}~\bibnamefont {Gupta}}, \bibinfo
  {author} {\bibfnamefont {M.~W.}\ \bibnamefont {Zwierlein}}, \bibinfo {author}
  {\bibfnamefont {A.}~\bibnamefont {G\"orlitz}},\ and\ \bibinfo {author}
  {\bibfnamefont {W.}~\bibnamefont {Ketterle}},\ }\bibfield  {title} {\bibinfo
  {title} {Two-species mixture of quantum degenerate bose and fermi gases},\
  }\href {https://doi.org/10.1103/PhysRevLett.88.160401} {\bibfield  {journal}
  {\bibinfo  {journal} {Phys. Rev. Lett.}\ }\textbf {\bibinfo {volume} {88}},\
  \bibinfo {pages} {160401} (\bibinfo {year} {2002})}\BibitemShut {NoStop}%
\bibitem [{\citenamefont {Goldwin}\ \emph {et~al.}(2004)\citenamefont
  {Goldwin}, \citenamefont {Inouye}, \citenamefont {Olsen}, \citenamefont
  {Newman}, \citenamefont {DePaola},\ and\ \citenamefont
  {Jin}}]{Goldwin2004PhysRevA.70.021601}%
  \BibitemOpen
  \bibfield  {author} {\bibinfo {author} {\bibfnamefont {J.}~\bibnamefont
  {Goldwin}}, \bibinfo {author} {\bibfnamefont {S.}~\bibnamefont {Inouye}},
  \bibinfo {author} {\bibfnamefont {M.~L.}\ \bibnamefont {Olsen}}, \bibinfo
  {author} {\bibfnamefont {B.}~\bibnamefont {Newman}}, \bibinfo {author}
  {\bibfnamefont {B.~D.}\ \bibnamefont {DePaola}},\ and\ \bibinfo {author}
  {\bibfnamefont {D.~S.}\ \bibnamefont {Jin}},\ }\bibfield  {title} {\bibinfo
  {title} {Measurement of the interaction strength in a bose-fermi mixture with
  $^{87}\mathrm{Rb}$ and $^{40}\mathrm{K}$},\ }\href
  {https://doi.org/10.1103/PhysRevA.70.021601} {\bibfield  {journal} {\bibinfo
  {journal} {Phys. Rev. A}\ }\textbf {\bibinfo {volume} {70}},\ \bibinfo
  {pages} {021601} (\bibinfo {year} {2004})}\BibitemShut {NoStop}%
\bibitem [{\citenamefont {Marco}\ \emph {et~al.}(2019)\citenamefont {Marco},
  \citenamefont {Valtolina}, \citenamefont {Matsuda}, \citenamefont {Tobias},
  \citenamefont {Covey},\ and\ \citenamefont {Ye}}]{Marco2019Science.363.853}%
  \BibitemOpen
  \bibfield  {author} {\bibinfo {author} {\bibfnamefont {L.~D.}\ \bibnamefont
  {Marco}}, \bibinfo {author} {\bibfnamefont {G.}~\bibnamefont {Valtolina}},
  \bibinfo {author} {\bibfnamefont {K.}~\bibnamefont {Matsuda}}, \bibinfo
  {author} {\bibfnamefont {W.~G.}\ \bibnamefont {Tobias}}, \bibinfo {author}
  {\bibfnamefont {J.~P.}\ \bibnamefont {Covey}},\ and\ \bibinfo {author}
  {\bibfnamefont {J.}~\bibnamefont {Ye}},\ }\bibfield  {title} {\bibinfo
  {title} {A degenerate fermi gas of polar molecules},\ }\href
  {https://doi.org/10.1126/science.aau7230} {\bibfield  {journal} {\bibinfo
  {journal} {Science}\ }\textbf {\bibinfo {volume} {363}},\ \bibinfo {pages}
  {853} (\bibinfo {year} {2019})},\ \Eprint
  {https://arxiv.org/abs/https://www.science.org/doi/pdf/10.1126/science.aau7230}
  {https://www.science.org/doi/pdf/10.1126/science.aau7230} \BibitemShut
  {NoStop}%
\bibitem [{\citenamefont {Fukuhara}\ \emph {et~al.}(2009)\citenamefont
  {Fukuhara}, \citenamefont {Sugawa}, \citenamefont {Takasu},\ and\
  \citenamefont {Takahashi}}]{Fukuhara2009PhysRevA.79.021601}%
  \BibitemOpen
  \bibfield  {author} {\bibinfo {author} {\bibfnamefont {T.}~\bibnamefont
  {Fukuhara}}, \bibinfo {author} {\bibfnamefont {S.}~\bibnamefont {Sugawa}},
  \bibinfo {author} {\bibfnamefont {Y.}~\bibnamefont {Takasu}},\ and\ \bibinfo
  {author} {\bibfnamefont {Y.}~\bibnamefont {Takahashi}},\ }\bibfield  {title}
  {\bibinfo {title} {All-optical formation of quantum degenerate mixtures},\
  }\href {https://doi.org/10.1103/PhysRevA.79.021601} {\bibfield  {journal}
  {\bibinfo  {journal} {Phys. Rev. A}\ }\textbf {\bibinfo {volume} {79}},\
  \bibinfo {pages} {021601} (\bibinfo {year} {2009})}\BibitemShut {NoStop}%
\bibitem [{\citenamefont {Tey}\ \emph {et~al.}(2010)\citenamefont {Tey},
  \citenamefont {Stellmer}, \citenamefont {Grimm},\ and\ \citenamefont
  {Schreck}}]{Tey2010PhysRevA.82.011608}%
  \BibitemOpen
  \bibfield  {author} {\bibinfo {author} {\bibfnamefont {M.~K.}\ \bibnamefont
  {Tey}}, \bibinfo {author} {\bibfnamefont {S.}~\bibnamefont {Stellmer}},
  \bibinfo {author} {\bibfnamefont {R.}~\bibnamefont {Grimm}},\ and\ \bibinfo
  {author} {\bibfnamefont {F.}~\bibnamefont {Schreck}},\ }\bibfield  {title}
  {\bibinfo {title} {Double-degenerate bose-fermi mixture of strontium},\
  }\href {https://doi.org/10.1103/PhysRevA.82.011608} {\bibfield  {journal}
  {\bibinfo  {journal} {Phys. Rev. A}\ }\textbf {\bibinfo {volume} {82}},\
  \bibinfo {pages} {011608} (\bibinfo {year} {2010})}\BibitemShut {NoStop}%
\bibitem [{\citenamefont {Wu}\ \emph {et~al.}(2011)\citenamefont {Wu},
  \citenamefont {Santiago}, \citenamefont {Park}, \citenamefont {Ahmadi},\ and\
  \citenamefont {Zwierlein}}]{Wu2011PhysRevA.84.011601}%
  \BibitemOpen
  \bibfield  {author} {\bibinfo {author} {\bibfnamefont {C.-H.}\ \bibnamefont
  {Wu}}, \bibinfo {author} {\bibfnamefont {I.}~\bibnamefont {Santiago}},
  \bibinfo {author} {\bibfnamefont {J.~W.}\ \bibnamefont {Park}}, \bibinfo
  {author} {\bibfnamefont {P.}~\bibnamefont {Ahmadi}},\ and\ \bibinfo {author}
  {\bibfnamefont {M.~W.}\ \bibnamefont {Zwierlein}},\ }\bibfield  {title}
  {\bibinfo {title} {Strongly interacting isotopic bose-fermi mixture immersed
  in a fermi sea},\ }\href {https://doi.org/10.1103/PhysRevA.84.011601}
  {\bibfield  {journal} {\bibinfo  {journal} {Phys. Rev. A}\ }\textbf {\bibinfo
  {volume} {84}},\ \bibinfo {pages} {011601} (\bibinfo {year}
  {2011})}\BibitemShut {NoStop}%
\bibitem [{\citenamefont {Ferrier-Barbut}\ \emph {et~al.}(2014)\citenamefont
  {Ferrier-Barbut}, \citenamefont {Delehaye}, \citenamefont {Laurent},
  \citenamefont {Grier}, \citenamefont {Pierce}, \citenamefont {Rem},
  \citenamefont {Chevy},\ and\ \citenamefont
  {Salomon}}]{Ferrier2014Science.345.1035}%
  \BibitemOpen
  \bibfield  {author} {\bibinfo {author} {\bibfnamefont {I.}~\bibnamefont
  {Ferrier-Barbut}}, \bibinfo {author} {\bibfnamefont {M.}~\bibnamefont
  {Delehaye}}, \bibinfo {author} {\bibfnamefont {S.}~\bibnamefont {Laurent}},
  \bibinfo {author} {\bibfnamefont {A.~T.}\ \bibnamefont {Grier}}, \bibinfo
  {author} {\bibfnamefont {M.}~\bibnamefont {Pierce}}, \bibinfo {author}
  {\bibfnamefont {B.~S.}\ \bibnamefont {Rem}}, \bibinfo {author} {\bibfnamefont
  {F.}~\bibnamefont {Chevy}},\ and\ \bibinfo {author} {\bibfnamefont
  {C.}~\bibnamefont {Salomon}},\ }\bibfield  {title} {\bibinfo {title} {A
  mixture of bose and fermi superfluids},\ }\href
  {https://doi.org/10.1126/science.1255380} {\bibfield  {journal} {\bibinfo
  {journal} {Science}\ }\textbf {\bibinfo {volume} {345}},\ \bibinfo {pages}
  {1035} (\bibinfo {year} {2014})}\BibitemShut {NoStop}%
\bibitem [{\citenamefont {Ikemachi}\ \emph {et~al.}(2016)\citenamefont
  {Ikemachi}, \citenamefont {Ito}, \citenamefont {Aratake}, \citenamefont
  {Chen}, \citenamefont {Koashi}, \citenamefont {Kuwata-Gonokami},\ and\
  \citenamefont {Horikoshi}}]{ikemachi2016all}%
  \BibitemOpen
  \bibfield  {author} {\bibinfo {author} {\bibfnamefont {T.}~\bibnamefont
  {Ikemachi}}, \bibinfo {author} {\bibfnamefont {A.}~\bibnamefont {Ito}},
  \bibinfo {author} {\bibfnamefont {Y.}~\bibnamefont {Aratake}}, \bibinfo
  {author} {\bibfnamefont {Y.}~\bibnamefont {Chen}}, \bibinfo {author}
  {\bibfnamefont {M.}~\bibnamefont {Koashi}}, \bibinfo {author} {\bibfnamefont
  {M.}~\bibnamefont {Kuwata-Gonokami}},\ and\ \bibinfo {author} {\bibfnamefont
  {M.}~\bibnamefont {Horikoshi}},\ }\bibfield  {title} {\bibinfo {title}
  {All-optical production of dual bose--einstein condensates of paired fermions
  and bosons with 6li and 7li},\ } {\bibfield  {journal} {\bibinfo
   {journal} {J. Phys. B: At. Mol. Opt. Phys.}\ }\textbf {\bibinfo {volume}
  {50}},\ \bibinfo {pages} {01LT01} (\bibinfo {year} {2016})}\BibitemShut
  {NoStop}%
\bibitem [{\citenamefont {Park}\ \emph {et~al.}(2012)\citenamefont {Park},
  \citenamefont {Wu}, \citenamefont {Santiago}, \citenamefont {Tiecke},
  \citenamefont {Will}, \citenamefont {Ahmadi},\ and\ \citenamefont
  {Zwierlein}}]{Park2012PhysRevA.85.051602}%
  \BibitemOpen
  \bibfield  {author} {\bibinfo {author} {\bibfnamefont {J.~W.}\ \bibnamefont
  {Park}}, \bibinfo {author} {\bibfnamefont {C.-H.}\ \bibnamefont {Wu}},
  \bibinfo {author} {\bibfnamefont {I.}~\bibnamefont {Santiago}}, \bibinfo
  {author} {\bibfnamefont {T.~G.}\ \bibnamefont {Tiecke}}, \bibinfo {author}
  {\bibfnamefont {S.}~\bibnamefont {Will}}, \bibinfo {author} {\bibfnamefont
  {P.}~\bibnamefont {Ahmadi}},\ and\ \bibinfo {author} {\bibfnamefont {M.~W.}\
  \bibnamefont {Zwierlein}},\ }\bibfield  {title} {\bibinfo {title} {Quantum
  degenerate bose-fermi mixture of chemically different atomic species with
  widely tunable interactions},\ }\href
  {https://doi.org/10.1103/PhysRevA.85.051602} {\bibfield  {journal} {\bibinfo
  {journal} {Phys. Rev. A}\ }\textbf {\bibinfo {volume} {85}},\ \bibinfo
  {pages} {051602} (\bibinfo {year} {2012})}\BibitemShut {NoStop}%
\bibitem [{\citenamefont {Duda}\ \emph
  {et~al.}(2023{\natexlab{a}})\citenamefont {Duda}, \citenamefont {Chen},
  \citenamefont {Schindewolf}, \citenamefont {Bause}, \citenamefont {von
  Milczewski}, \citenamefont {Schmidt}, \citenamefont {Bloch},\ and\
  \citenamefont {Luo}}]{duda2021transition}%
  \BibitemOpen
  \bibfield  {author} {\bibinfo {author} {\bibfnamefont {M.}~\bibnamefont
  {Duda}}, \bibinfo {author} {\bibfnamefont {X.-Y.}\ \bibnamefont {Chen}},
  \bibinfo {author} {\bibfnamefont {A.}~\bibnamefont {Schindewolf}}, \bibinfo
  {author} {\bibfnamefont {R.}~\bibnamefont {Bause}}, \bibinfo {author}
  {\bibfnamefont {J.}~\bibnamefont {von Milczewski}}, \bibinfo {author}
  {\bibfnamefont {R.}~\bibnamefont {Schmidt}}, \bibinfo {author} {\bibfnamefont
  {I.}~\bibnamefont {Bloch}},\ and\ \bibinfo {author} {\bibfnamefont {X.-Y.}\
  \bibnamefont {Luo}},\ }\bibfield  {title} {\bibinfo {title} {Transition from
  a polaronic condensate to a degenerate fermi gas of heteronuclear
  molecules},\ }\href {https://doi.org/10.1038/s41567-023-01948-1} {\bibfield
  {journal} {\bibinfo  {journal} {Nat. Phys.}\ }\textbf {\bibinfo {volume}
  {19}},\ \bibinfo {pages} {720} (\bibinfo {year}
  {2023}{\natexlab{a}})}\BibitemShut {NoStop}%
\bibitem [{\citenamefont {Duda}\ \emph
  {et~al.}(2023{\natexlab{b}})\citenamefont {Duda}, \citenamefont {Chen},
  \citenamefont {Bause}, \citenamefont {Schindewolf}, \citenamefont {Bloch},\
  and\ \citenamefont {Luo}}]{duda2022longlived}%
  \BibitemOpen
  \bibfield  {author} {\bibinfo {author} {\bibfnamefont {M.}~\bibnamefont
  {Duda}}, \bibinfo {author} {\bibfnamefont {X.-Y.}\ \bibnamefont {Chen}},
  \bibinfo {author} {\bibfnamefont {R.}~\bibnamefont {Bause}}, \bibinfo
  {author} {\bibfnamefont {A.}~\bibnamefont {Schindewolf}}, \bibinfo {author}
  {\bibfnamefont {I.}~\bibnamefont {Bloch}},\ and\ \bibinfo {author}
  {\bibfnamefont {X.-Y.}\ \bibnamefont {Luo}},\ }\bibfield  {title} {\bibinfo
  {title} {Long-lived fermionic feshbach molecules with tunable $p$-wave
  interactions},\ }{\bibfield  {journal} {\bibinfo  {journal}
  {Phys. Rev. A}\ }\textbf {\bibinfo {volume} {107}},\ \bibinfo {pages}
  {053322} (\bibinfo {year} {2023}{\natexlab{b}})}\BibitemShut {NoStop}%
\bibitem [{\citenamefont {Chen}\ \emph {et~al.}(2023)\citenamefont {Chen},
  \citenamefont {Schindewolf}, \citenamefont {Eppelt}, \citenamefont {Bause},
  \citenamefont {Duda}, \citenamefont {Biswas}, \citenamefont {Karman},
  \citenamefont {Hilker}, \citenamefont {Bloch},\ and\ \citenamefont
  {Luo}}]{chen2023field}%
  \BibitemOpen
  \bibfield  {author} {\bibinfo {author} {\bibfnamefont {X.-Y.}\ \bibnamefont
  {Chen}}, \bibinfo {author} {\bibfnamefont {A.}~\bibnamefont {Schindewolf}},
  \bibinfo {author} {\bibfnamefont {S.}~\bibnamefont {Eppelt}}, \bibinfo
  {author} {\bibfnamefont {R.}~\bibnamefont {Bause}}, \bibinfo {author}
  {\bibfnamefont {M.}~\bibnamefont {Duda}}, \bibinfo {author} {\bibfnamefont
  {S.}~\bibnamefont {Biswas}}, \bibinfo {author} {\bibfnamefont
  {T.}~\bibnamefont {Karman}}, \bibinfo {author} {\bibfnamefont
  {T.}~\bibnamefont {Hilker}}, \bibinfo {author} {\bibfnamefont
  {I.}~\bibnamefont {Bloch}},\ and\ \bibinfo {author} {\bibfnamefont {X.-Y.}\
  \bibnamefont {Luo}},\ }\bibfield  {title} {\bibinfo {title} {Field-linked
  resonances of polar molecules},\ }{\bibfield  {journal}
  {\bibinfo  {journal} {Nature}\ }\textbf {\bibinfo {volume} {614}},\ \bibinfo
  {pages} {59} (\bibinfo {year} {2023})}\BibitemShut {NoStop}%
\bibitem [{\citenamefont {Baranov}(2008)}]{BARANOV200871}%
  \BibitemOpen
  \bibfield  {author} {\bibinfo {author} {\bibfnamefont {M.}~\bibnamefont
  {Baranov}},\ }\bibfield  {title} {\bibinfo {title} {Theoretical progress in
  many-body physics with ultracold dipolar gases},\ }\href
  {https://doi.org/https://doi.org/10.1016/j.physrep.2008.04.007} {\bibfield
  {journal} {\bibinfo  {journal} {Physics Reports}\ }\textbf {\bibinfo {volume}
  {464}},\ \bibinfo {pages} {71} (\bibinfo {year} {2008})}\BibitemShut
  {NoStop}%
\bibitem [{\citenamefont {Kinnunen}\ \emph {et~al.}(2018)\citenamefont
  {Kinnunen}, \citenamefont {Wu},\ and\ \citenamefont
  {Bruun}}]{Kinnunen2018PhysRevLett.121.253402}%
  \BibitemOpen
  \bibfield  {author} {\bibinfo {author} {\bibfnamefont {J.~J.}\ \bibnamefont
  {Kinnunen}}, \bibinfo {author} {\bibfnamefont {Z.}~\bibnamefont {Wu}},\ and\
  \bibinfo {author} {\bibfnamefont {G.~M.}\ \bibnamefont {Bruun}},\ }\bibfield
  {title} {\bibinfo {title} {Induced $p$-wave pairing in bose-fermi mixtures},\
  }\href {https://doi.org/10.1103/PhysRevLett.121.253402} {\bibfield  {journal}
  {\bibinfo  {journal} {Phys. Rev. Lett.}\ }\textbf {\bibinfo {volume} {121}},\
  \bibinfo {pages} {253402} (\bibinfo {year} {2018})}\BibitemShut {NoStop}%
\bibitem [{\citenamefont {Fujimoto}\ \emph {et~al.}(2020)\citenamefont
  {Fujimoto}, \citenamefont {Fukushima},\ and\ \citenamefont
  {Weise}}]{Fujimoto:2019sxg}%
  \BibitemOpen
  \bibfield  {author} {\bibinfo {author} {\bibfnamefont {Y.}~\bibnamefont
  {Fujimoto}}, \bibinfo {author} {\bibfnamefont {K.}~\bibnamefont
  {Fukushima}},\ and\ \bibinfo {author} {\bibfnamefont {W.}~\bibnamefont
  {Weise}},\ }\bibfield  {title} {\bibinfo {title} {{Continuity from neutron
  matter to two-flavor quark matter with $^1 S_0$ and $^3 P_2$
  superfluidity}},\ }\href {https://doi.org/10.1103/PhysRevD.101.094009}
  {\bibfield  {journal} {\bibinfo  {journal} {Phys. Rev. D}\ }\textbf {\bibinfo
  {volume} {101}},\ \bibinfo {pages} {094009} (\bibinfo {year}
  {2020})}\BibitemShut {NoStop}%
\bibitem [{\citenamefont {Sch\"afer}\ and\ \citenamefont
  {Wilczek}(1999)}]{Schafer1999PhysRevLett.82.3956}%
  \BibitemOpen
  \bibfield  {author} {\bibinfo {author} {\bibfnamefont {T.}~\bibnamefont
  {Sch\"afer}}\ and\ \bibinfo {author} {\bibfnamefont {F.}~\bibnamefont
  {Wilczek}},\ }\bibfield  {title} {\bibinfo {title} {Continuity of quark and
  hadron matter},\ }\href {https://doi.org/10.1103/PhysRevLett.82.3956}
  {\bibfield  {journal} {\bibinfo  {journal} {Phys. Rev. Lett.}\ }\textbf
  {\bibinfo {volume} {82}},\ \bibinfo {pages} {3956} (\bibinfo {year}
  {1999})}\BibitemShut {NoStop}%
\bibitem [{\citenamefont {Hatsuda}\ \emph {et~al.}(2006)\citenamefont
  {Hatsuda}, \citenamefont {Tachibana}, \citenamefont {Yamamoto},\ and\
  \citenamefont {Baym}}]{Hatsuda2006PhysRevLett.97.122001}%
  \BibitemOpen
  \bibfield  {author} {\bibinfo {author} {\bibfnamefont {T.}~\bibnamefont
  {Hatsuda}}, \bibinfo {author} {\bibfnamefont {M.}~\bibnamefont {Tachibana}},
  \bibinfo {author} {\bibfnamefont {N.}~\bibnamefont {Yamamoto}},\ and\
  \bibinfo {author} {\bibfnamefont {G.}~\bibnamefont {Baym}},\ }\bibfield
  {title} {\bibinfo {title} {New critical point induced by the axial anomaly in
  dense qcd},\ }\href {https://doi.org/10.1103/PhysRevLett.97.122001}
  {\bibfield  {journal} {\bibinfo  {journal} {Phys. Rev. Lett.}\ }\textbf
  {\bibinfo {volume} {97}},\ \bibinfo {pages} {122001} (\bibinfo {year}
  {2006})}\BibitemShut {NoStop}%
\bibitem [{\citenamefont {Maeda}\ \emph {et~al.}(2009)\citenamefont {Maeda},
  \citenamefont {Baym},\ and\ \citenamefont
  {Hatsuda}}]{Maeda2009PhysRevLett.103.085301}%
  \BibitemOpen
  \bibfield  {author} {\bibinfo {author} {\bibfnamefont {K.}~\bibnamefont
  {Maeda}}, \bibinfo {author} {\bibfnamefont {G.}~\bibnamefont {Baym}},\ and\
  \bibinfo {author} {\bibfnamefont {T.}~\bibnamefont {Hatsuda}},\ }\bibfield
  {title} {\bibinfo {title} {Simulating dense qcd matter with ultracold atomic
  boson-fermion mixtures},\ }\href
  {https://doi.org/10.1103/PhysRevLett.103.085301} {\bibfield  {journal}
  {\bibinfo  {journal} {Phys. Rev. Lett.}\ }\textbf {\bibinfo {volume} {103}},\
  \bibinfo {pages} {085301} (\bibinfo {year} {2009})}\BibitemShut {NoStop}%
\bibitem [{\citenamefont {Baym}\ \emph {et~al.}(2018)\citenamefont {Baym},
  \citenamefont {Hatsuda}, \citenamefont {Kojo}, \citenamefont {Powell},
  \citenamefont {Song},\ and\ \citenamefont
  {Takatsuka}}]{Baym2018Rep.Prog.Phys.81.056902}%
  \BibitemOpen
  \bibfield  {author} {\bibinfo {author} {\bibfnamefont {G.}~\bibnamefont
  {Baym}}, \bibinfo {author} {\bibfnamefont {T.}~\bibnamefont {Hatsuda}},
  \bibinfo {author} {\bibfnamefont {T.}~\bibnamefont {Kojo}}, \bibinfo {author}
  {\bibfnamefont {P.~D.}\ \bibnamefont {Powell}}, \bibinfo {author}
  {\bibfnamefont {Y.}~\bibnamefont {Song}},\ and\ \bibinfo {author}
  {\bibfnamefont {T.}~\bibnamefont {Takatsuka}},\ }\bibfield  {title} {\bibinfo
  {title} {From hadrons to quarks in neutron stars: a review},\ }
  {\bibfield  {journal} {\bibinfo  {journal} {Rep. Prog. Phys.}\ }\textbf
  {\bibinfo {volume} {81}},\ \bibinfo {pages} {056902} (\bibinfo {year}
  {2018})}\BibitemShut {NoStop}%
\bibitem [{\citenamefont {Chin}\ \emph {et~al.}(2010)\citenamefont {Chin},
  \citenamefont {Grimm}, \citenamefont {Julienne},\ and\ \citenamefont
  {Tiesinga}}]{Chin2010RevModPhys.82.1225}%
  \BibitemOpen
  \bibfield  {author} {\bibinfo {author} {\bibfnamefont {C.}~\bibnamefont
  {Chin}}, \bibinfo {author} {\bibfnamefont {R.}~\bibnamefont {Grimm}},
  \bibinfo {author} {\bibfnamefont {P.}~\bibnamefont {Julienne}},\ and\
  \bibinfo {author} {\bibfnamefont {E.}~\bibnamefont {Tiesinga}},\ }\bibfield
  {title} {\bibinfo {title} {Feshbach resonances in ultracold gases},\ }\href
  {https://doi.org/10.1103/RevModPhys.82.1225} {\bibfield  {journal} {\bibinfo
  {journal} {Rev. Mod. Phys.}\ }\textbf {\bibinfo {volume} {82}},\ \bibinfo
  {pages} {1225} (\bibinfo {year} {2010})}\BibitemShut {NoStop}%
\bibitem [{\citenamefont {Viverit}\ \emph {et~al.}(2000)\citenamefont
  {Viverit}, \citenamefont {Pethick},\ and\ \citenamefont
  {Smith}}]{Viverit2000Phys.Rev.A61.053605}%
  \BibitemOpen
  \bibfield  {author} {\bibinfo {author} {\bibfnamefont {L.}~\bibnamefont
  {Viverit}}, \bibinfo {author} {\bibfnamefont {C.~J.}\ \bibnamefont
  {Pethick}},\ and\ \bibinfo {author} {\bibfnamefont {H.}~\bibnamefont
  {Smith}},\ }\bibfield  {title} {\bibinfo {title} {Zero-temperature phase
  diagram of binary boson-fermion mixtures},\ }\href
  {https://doi.org/10.1103/PhysRevA.61.053605} {\bibfield  {journal} {\bibinfo
  {journal} {Phys. Rev. A}\ }\textbf {\bibinfo {volume} {61}},\ \bibinfo
  {pages} {053605} (\bibinfo {year} {2000})}\BibitemShut {NoStop}%
\bibitem [{\citenamefont {Viverit}\ and\ \citenamefont
  {Giorgini}(2002)}]{Viverit2002Phys.Rev.A66.063604}%
  \BibitemOpen
  \bibfield  {author} {\bibinfo {author} {\bibfnamefont {L.}~\bibnamefont
  {Viverit}}\ and\ \bibinfo {author} {\bibfnamefont {S.}~\bibnamefont
  {Giorgini}},\ }\bibfield  {title} {\bibinfo {title} {Ground-state properties
  of a dilute bose-fermi mixture},\ }\href
  {https://doi.org/10.1103/PhysRevA.66.063604} {\bibfield  {journal} {\bibinfo
  {journal} {Phys. Rev. A}\ }\textbf {\bibinfo {volume} {66}},\ \bibinfo
  {pages} {063604} (\bibinfo {year} {2002})}\BibitemShut {NoStop}%
\bibitem [{\citenamefont {Efremov}\ and\ \citenamefont
  {Viverit}(2002)}]{Efremov2002Phys.Rev.B65.134519}%
  \BibitemOpen
  \bibfield  {author} {\bibinfo {author} {\bibfnamefont {D.~V.}\ \bibnamefont
  {Efremov}}\ and\ \bibinfo {author} {\bibfnamefont {L.}~\bibnamefont
  {Viverit}},\ }\bibfield  {title} {\bibinfo {title} {$p$-wave cooper pairing
  of fermions in mixtures of dilute fermi and bose gases},\ }\href
  {https://doi.org/10.1103/PhysRevB.65.134519} {\bibfield  {journal} {\bibinfo
  {journal} {Phys. Rev. B}\ }\textbf {\bibinfo {volume} {65}},\ \bibinfo
  {pages} {134519} (\bibinfo {year} {2002})}\BibitemShut {NoStop}%
\bibitem [{\citenamefont {Fratini}\ and\ \citenamefont
  {Pieri}(2010)}]{Fratini2010Phys.Rev.A81.051605}%
  \BibitemOpen
  \bibfield  {author} {\bibinfo {author} {\bibfnamefont {E.}~\bibnamefont
  {Fratini}}\ and\ \bibinfo {author} {\bibfnamefont {P.}~\bibnamefont
  {Pieri}},\ }\bibfield  {title} {\bibinfo {title} {Pairing and condensation in
  a resonant bose-fermi mixture},\ }\href
  {https://doi.org/10.1103/PhysRevA.81.051605} {\bibfield  {journal} {\bibinfo
  {journal} {Phys. Rev. A}\ }\textbf {\bibinfo {volume} {81}},\ \bibinfo
  {pages} {051605} (\bibinfo {year} {2010})}\BibitemShut {NoStop}%
\bibitem [{\citenamefont {Guidini}\ \emph {et~al.}(2015)\citenamefont
  {Guidini}, \citenamefont {Bertaina}, \citenamefont {Galli},\ and\
  \citenamefont {Pieri}}]{Guidini2015Phys.Rev.A91.023603}%
  \BibitemOpen
  \bibfield  {author} {\bibinfo {author} {\bibfnamefont {A.}~\bibnamefont
  {Guidini}}, \bibinfo {author} {\bibfnamefont {G.}~\bibnamefont {Bertaina}},
  \bibinfo {author} {\bibfnamefont {D.~E.}\ \bibnamefont {Galli}},\ and\
  \bibinfo {author} {\bibfnamefont {P.}~\bibnamefont {Pieri}},\ }\bibfield
  {title} {\bibinfo {title} {Condensed phase of bose-fermi mixtures with a
  pairing interaction},\ }\href {https://doi.org/10.1103/PhysRevA.91.023603}
  {\bibfield  {journal} {\bibinfo  {journal} {Phys. Rev. A}\ }\textbf {\bibinfo
  {volume} {91}},\ \bibinfo {pages} {023603} (\bibinfo {year}
  {2015})}\BibitemShut {NoStop}%
\bibitem [{\citenamefont {Suhl}\ \emph {et~al.}(1959)\citenamefont {Suhl},
  \citenamefont {Matthias},\ and\ \citenamefont
  {Walker}}]{Suhl1959PhysRevLett.3.552}%
  \BibitemOpen
  \bibfield  {author} {\bibinfo {author} {\bibfnamefont {H.}~\bibnamefont
  {Suhl}}, \bibinfo {author} {\bibfnamefont {B.~T.}\ \bibnamefont {Matthias}},\
  and\ \bibinfo {author} {\bibfnamefont {L.~R.}\ \bibnamefont {Walker}},\
  }\bibfield  {title} {\bibinfo {title} {Bardeen-cooper-schrieffer theory of
  superconductivity in the case of overlapping bands},\ }\href
  {https://doi.org/10.1103/PhysRevLett.3.552} {\bibfield  {journal} {\bibinfo
  {journal} {Phys. Rev. Lett.}\ }\textbf {\bibinfo {volume} {3}},\ \bibinfo
  {pages} {552} (\bibinfo {year} {1959})}\BibitemShut {NoStop}%
\bibitem [{\citenamefont {Kondo}(1963)}]{Kondo196310.1143/PTP.29.1}%
  \BibitemOpen
  \bibfield  {author} {\bibinfo {author} {\bibfnamefont {J.}~\bibnamefont
  {Kondo}},\ }\bibfield  {title} {\bibinfo {title} {Superconductivity in
  transition metals},\ }\href {https://doi.org/10.1143/PTP.29.1} {\bibfield
  {journal} {\bibinfo  {journal} {Prog. Theor. Phys.}\ }\textbf {\bibinfo
  {volume} {29}},\ \bibinfo {pages} {1} (\bibinfo {year} {1963})}\BibitemShut
  {NoStop}%
\bibitem [{\citenamefont {Aoki}(2020)}]{aoki2020theoretical}%
  \BibitemOpen
  \bibfield  {author} {\bibinfo {author} {\bibfnamefont {H.}~\bibnamefont
  {Aoki}},\ }\bibfield  {title} {\bibinfo {title} {Theoretical possibilities
  for flat band superconductivity},\ }{\bibfield  {journal}
  {\bibinfo  {journal} {J. Supercond. Nov. Magn.}\ }\textbf {\bibinfo {volume}
  {33}},\ \bibinfo {pages} {2341} (\bibinfo {year} {2020})}\BibitemShut
  {NoStop}%
\bibitem [{\citenamefont {Ohashi}\ \emph {et~al.}(2020)\citenamefont {Ohashi},
  \citenamefont {Tajima},\ and\ \citenamefont {{van Wyk}}}]{OHASHI2020103739}%
  \BibitemOpen
  \bibfield  {author} {\bibinfo {author} {\bibfnamefont {Y.}~\bibnamefont
  {Ohashi}}, \bibinfo {author} {\bibfnamefont {H.}~\bibnamefont {Tajima}},\
  and\ \bibinfo {author} {\bibfnamefont {P.}~\bibnamefont {{van Wyk}}},\
  }\bibfield  {title} {\bibinfo {title} {Bcs–bec crossover in cold atomic and
  in nuclear systems},\ }\href
  {https://doi.org/https://doi.org/10.1016/j.ppnp.2019.103739} {\bibfield
  {journal} {\bibinfo  {journal} {Prog. Part. Nucl. Phys.}\ }\textbf {\bibinfo
  {volume} {111}},\ \bibinfo {pages} {103739} (\bibinfo {year}
  {2020})}\BibitemShut {NoStop}%
\bibitem [{\citenamefont {Goldman}\ \emph {et~al.}(2014)\citenamefont
  {Goldman}, \citenamefont {Juzeli\={u}nas}, \citenamefont {\"{O}hberg},\ and\
  \citenamefont {Spielman}}]{Goldman2014}%
  \BibitemOpen
  \bibfield  {author} {\bibinfo {author} {\bibfnamefont {N.}~\bibnamefont
  {Goldman}}, \bibinfo {author} {\bibfnamefont {G.}~\bibnamefont
  {Juzeli\={u}nas}}, \bibinfo {author} {\bibfnamefont {P.}~\bibnamefont
  {\"{O}hberg}},\ and\ \bibinfo {author} {\bibfnamefont {I.~B.}\ \bibnamefont
  {Spielman}},\ }\bibfield  {title} {\bibinfo {title} {Light-induced gauge
  fields for ultracold atoms},\ } {\bibfield  {journal} {\bibinfo
  {journal} {Rep. Prog. Phys.}\ }\textbf {\bibinfo {volume} {77}},\ \bibinfo
  {pages} {126401} (\bibinfo {year} {2014})}\BibitemShut {NoStop}%
\bibitem [{\citenamefont {Chow}(1968)}]{Chow1968PhysRev.172.467}%
  \BibitemOpen
  \bibfield  {author} {\bibinfo {author} {\bibfnamefont {W.~S.}\ \bibnamefont
  {Chow}},\ }\bibfield  {title} {\bibinfo {title} {Theory of superconductors
  with overlapping bands in the presence of nonmagnetic impurities},\ }\href
  {https://doi.org/10.1103/PhysRev.172.467} {\bibfield  {journal} {\bibinfo
  {journal} {Phys. Rev.}\ }\textbf {\bibinfo {volume} {172}},\ \bibinfo {pages}
  {467} (\bibinfo {year} {1968})}\BibitemShut {NoStop}%
\bibitem [{\citenamefont {Camacho-Guardian}\ \emph {et~al.}(2018)\citenamefont
  {Camacho-Guardian}, \citenamefont {Ardila}, \citenamefont {Pohl},\ and\
  \citenamefont {Bruun}}]{CamachoGuardian2018Phys.Rev.Lett.121.013401}%
  \BibitemOpen
  \bibfield  {author} {\bibinfo {author} {\bibfnamefont {A.}~\bibnamefont
  {Camacho-Guardian}}, \bibinfo {author} {\bibfnamefont {L.~P.}\ \bibnamefont
  {Ardila}}, \bibinfo {author} {\bibfnamefont {T.}~\bibnamefont {Pohl}},\ and\
  \bibinfo {author} {\bibfnamefont {G.}~\bibnamefont {Bruun}},\ }\bibfield
  {title} {\bibinfo {title} {Bipolarons in a bose-einstein condensate},\ }\href
  {https://doi.org/10.1103/physrevlett.121.013401} {\bibfield  {journal}
  {\bibinfo  {journal} {Phys. Rev. Lett.}\ }\textbf {\bibinfo {volume} {121}},\
  \bibinfo {pages} {013401} (\bibinfo {year} {2018})}\BibitemShut {NoStop}%
\bibitem [{\citenamefont {Camacho-Guardian}\ and\ \citenamefont
  {Bruun}(2018)}]{CamachoGuardian2018Phys.Rev.X.8.031042}%
  \BibitemOpen
  \bibfield  {author} {\bibinfo {author} {\bibfnamefont {A.}~\bibnamefont
  {Camacho-Guardian}}\ and\ \bibinfo {author} {\bibfnamefont {G.~M.}\
  \bibnamefont {Bruun}},\ }\bibfield  {title} {\bibinfo {title} {Landau
  effective interaction between quasiparticles in a bose-einstein condensate},\
  }\href {https://doi.org/10.1103/PhysRevX.8.031042} {\bibfield  {journal}
  {\bibinfo  {journal} {Phys. Rev. X}\ }\textbf {\bibinfo {volume} {8}},\
  \bibinfo {pages} {031042} (\bibinfo {year} {2018})}\BibitemShut {NoStop}%
\bibitem [{\citenamefont {Gubbels}\ and\ \citenamefont
  {Stoof}(2013)}]{GUBBELS2013255}%
  \BibitemOpen
  \bibfield  {author} {\bibinfo {author} {\bibfnamefont {K.}~\bibnamefont
  {Gubbels}}\ and\ \bibinfo {author} {\bibfnamefont {H.}~\bibnamefont
  {Stoof}},\ }\bibfield  {title} {\bibinfo {title} {Imbalanced fermi gases at
  unitarity},\ }\href
  {https://doi.org/https://doi.org/10.1016/j.physrep.2012.11.004} {\bibfield
  {journal} {\bibinfo  {journal} {Phys. Rep.}\ }\textbf {\bibinfo {volume}
  {525}},\ \bibinfo {pages} {255} (\bibinfo {year} {2013})},\ \bibinfo {note}
  {imbalanced Fermi Gases at Unitarity}\BibitemShut {NoStop}%
\bibitem [{\citenamefont {Ochi}\ \emph {et~al.}(2022)\citenamefont {Ochi},
  \citenamefont {Tajima}, \citenamefont {Iida},\ and\ \citenamefont
  {Aoki}}]{Ochi2022PhysRevResearch.4.013032}%
  \BibitemOpen
  \bibfield  {author} {\bibinfo {author} {\bibfnamefont {K.}~\bibnamefont
  {Ochi}}, \bibinfo {author} {\bibfnamefont {H.}~\bibnamefont {Tajima}},
  \bibinfo {author} {\bibfnamefont {K.}~\bibnamefont {Iida}},\ and\ \bibinfo
  {author} {\bibfnamefont {H.}~\bibnamefont {Aoki}},\ }\bibfield  {title}
  {\bibinfo {title} {Resonant pair-exchange scattering and bcs-bec crossover in
  a system composed of dispersive and heavy incipient bands: A feshbach
  analogy},\ }\href {https://doi.org/10.1103/PhysRevResearch.4.013032}
  {\bibfield  {journal} {\bibinfo  {journal} {Phys. Rev. Research}\ }\textbf
  {\bibinfo {volume} {4}},\ \bibinfo {pages} {013032} (\bibinfo {year}
  {2022})}\BibitemShut {NoStop}%
\bibitem [{\citenamefont {Fetter}\ and\ \citenamefont
  {Walecka}(2012)}]{fetter2012quantum}%
  \BibitemOpen
  \bibfield  {author} {\bibinfo {author} {\bibfnamefont {A.~L.}\ \bibnamefont
  {Fetter}}\ and\ \bibinfo {author} {\bibfnamefont {J.~D.}\ \bibnamefont
  {Walecka}},\ } {\emph {\bibinfo {title} {Quantum theory of
  many-particle systems}}}\ (\bibinfo  {publisher} {Courier Corporation},\
  \bibinfo {year} {2012})\BibitemShut {NoStop}%
\bibitem [{\citenamefont {Giorgini}(1998)}]{Giorgini1998PhysRevA.57.2949}%
  \BibitemOpen
  \bibfield  {author} {\bibinfo {author} {\bibfnamefont {S.}~\bibnamefont
  {Giorgini}},\ }\bibfield  {title} {\bibinfo {title} {Damping in dilute bose
  gases: A mean-field approach},\ }\href
  {https://doi.org/10.1103/PhysRevA.57.2949} {\bibfield  {journal} {\bibinfo
  {journal} {Phys. Rev. A}\ }\textbf {\bibinfo {volume} {57}},\ \bibinfo
  {pages} {2949} (\bibinfo {year} {1998})}\BibitemShut {NoStop}%
\bibitem [{\citenamefont {Ho}\ and\ \citenamefont
  {Diener}(2005)}]{Ho2005PhysRevLett.94.090402}%
  \BibitemOpen
  \bibfield  {author} {\bibinfo {author} {\bibfnamefont {T.-L.}\ \bibnamefont
  {Ho}}\ and\ \bibinfo {author} {\bibfnamefont {R.~B.}\ \bibnamefont
  {Diener}},\ }\bibfield  {title} {\bibinfo {title} {Fermion superfluids of
  nonzero orbital angular momentum near resonance},\ }\href
  {https://doi.org/10.1103/PhysRevLett.94.090402} {\bibfield  {journal}
  {\bibinfo  {journal} {Phys. Rev. Lett.}\ }\textbf {\bibinfo {volume} {94}},\
  \bibinfo {pages} {090402} (\bibinfo {year} {2005})}\BibitemShut {NoStop}%
\bibitem [{\citenamefont {Silaev}\ and\ \citenamefont
  {Babaev}(2012)}]{Mihail2012PhysRevB.85.134514}%
  \BibitemOpen
  \bibfield  {author} {\bibinfo {author} {\bibfnamefont {M.}~\bibnamefont
  {Silaev}}\ and\ \bibinfo {author} {\bibfnamefont {E.}~\bibnamefont
  {Babaev}},\ }\bibfield  {title} {\bibinfo {title} {Microscopic derivation of
  two-component ginzburg-landau model and conditions of its applicability in
  two-band systems},\ }\href {https://doi.org/10.1103/PhysRevB.85.134514}
  {\bibfield  {journal} {\bibinfo  {journal} {Phys. Rev. B}\ }\textbf {\bibinfo
  {volume} {85}},\ \bibinfo {pages} {134514} (\bibinfo {year}
  {2012})}\BibitemShut {NoStop}%
\bibitem [{\citenamefont {Iskin}\ and\ \citenamefont {S\'a~de
  Melo}(2006)}]{PhysRevB.74.144517}%
  \BibitemOpen
  \bibfield  {author} {\bibinfo {author} {\bibfnamefont {M.}~\bibnamefont
  {Iskin}}\ and\ \bibinfo {author} {\bibfnamefont {C.~A.~R.}\ \bibnamefont
  {S\'a~de Melo}},\ }\bibfield  {title} {\bibinfo {title} {Two-band
  superfluidity from the bcs to the bec limit},\ }\href
  {https://doi.org/10.1103/PhysRevB.74.144517} {\bibfield  {journal} {\bibinfo
  {journal} {Phys. Rev. B}\ }\textbf {\bibinfo {volume} {74}},\ \bibinfo
  {pages} {144517} (\bibinfo {year} {2006})}\BibitemShut {NoStop}%
\bibitem [{\citenamefont {Dalfovo}\ \emph {et~al.}(1999)\citenamefont
  {Dalfovo}, \citenamefont {Giorgini}, \citenamefont {Pitaevskii},\ and\
  \citenamefont {Stringari}}]{Dalfovo1999RevModPhys.71.463}%
  \BibitemOpen
  \bibfield  {author} {\bibinfo {author} {\bibfnamefont {F.}~\bibnamefont
  {Dalfovo}}, \bibinfo {author} {\bibfnamefont {S.}~\bibnamefont {Giorgini}},
  \bibinfo {author} {\bibfnamefont {L.~P.}\ \bibnamefont {Pitaevskii}},\ and\
  \bibinfo {author} {\bibfnamefont {S.}~\bibnamefont {Stringari}},\ }\bibfield
  {title} {\bibinfo {title} {Theory of bose-einstein condensation in trapped
  gases},\ }\href {https://doi.org/10.1103/RevModPhys.71.463} {\bibfield
  {journal} {\bibinfo  {journal} {Rev. Mod. Phys.}\ }\textbf {\bibinfo {volume}
  {71}},\ \bibinfo {pages} {463} (\bibinfo {year} {1999})}\BibitemShut
  {NoStop}%
\bibitem [{\citenamefont {Marchetti}\ \emph {et~al.}(2008)\citenamefont
  {Marchetti}, \citenamefont {Mathy}, \citenamefont {Huse},\ and\ \citenamefont
  {Parish}}]{Marchetti2008Phys.Rev.B78.134517}%
  \BibitemOpen
  \bibfield  {author} {\bibinfo {author} {\bibfnamefont {F.~M.}\ \bibnamefont
  {Marchetti}}, \bibinfo {author} {\bibfnamefont {C.~J.~M.}\ \bibnamefont
  {Mathy}}, \bibinfo {author} {\bibfnamefont {D.~A.}\ \bibnamefont {Huse}},\
  and\ \bibinfo {author} {\bibfnamefont {M.~M.}\ \bibnamefont {Parish}},\
  }\bibfield  {title} {\bibinfo {title} {Phase separation and collapse in
  bose-fermi mixtures with a feshbach resonance},\ }\href
  {https://doi.org/10.1103/PhysRevB.78.134517} {\bibfield  {journal} {\bibinfo
  {journal} {Phys. Rev. B}\ }\textbf {\bibinfo {volume} {78}},\ \bibinfo
  {pages} {134517} (\bibinfo {year} {2008})}\BibitemShut {NoStop}%
\bibitem [{\citenamefont {Nishida}(2012)}]{Nishida2012PhysRevLett.109.240401}%
  \BibitemOpen
  \bibfield  {author} {\bibinfo {author} {\bibfnamefont {Y.}~\bibnamefont
  {Nishida}},\ }\bibfield  {title} {\bibinfo {title} {New type of crossover
  physics in three-component fermi gases},\ }\href
  {https://doi.org/10.1103/PhysRevLett.109.240401} {\bibfield  {journal}
  {\bibinfo  {journal} {Phys. Rev. Lett.}\ }\textbf {\bibinfo {volume} {109}},\
  \bibinfo {pages} {240401} (\bibinfo {year} {2012})}\BibitemShut {NoStop}%
\bibitem [{\citenamefont {Tajima}(2018)}]{Tajima2018PhysRevA.97.043613}%
  \BibitemOpen
  \bibfield  {author} {\bibinfo {author} {\bibfnamefont {H.}~\bibnamefont
  {Tajima}},\ }\bibfield  {title} {\bibinfo {title} {Precursor of superfluidity
  in a strongly interacting fermi gas with negative effective range},\ }\href
  {https://doi.org/10.1103/PhysRevA.97.043613} {\bibfield  {journal} {\bibinfo
  {journal} {Phys. Rev. A}\ }\textbf {\bibinfo {volume} {97}},\ \bibinfo
  {pages} {043613} (\bibinfo {year} {2018})}\BibitemShut {NoStop}%
\bibitem [{\citenamefont {Wu}\ and\ \citenamefont
  {Thomas}(2012)}]{Wu2012PhysRevA.86.063625}%
  \BibitemOpen
  \bibfield  {author} {\bibinfo {author} {\bibfnamefont {H.}~\bibnamefont
  {Wu}}\ and\ \bibinfo {author} {\bibfnamefont {J.~E.}\ \bibnamefont
  {Thomas}},\ }\bibfield  {title} {\bibinfo {title} {Optical control of the
  scattering length and effective range for magnetically tunable feshbach
  resonances in ultracold gases},\ }\href
  {https://doi.org/10.1103/PhysRevA.86.063625} {\bibfield  {journal} {\bibinfo
  {journal} {Phys. Rev. A}\ }\textbf {\bibinfo {volume} {86}},\ \bibinfo
  {pages} {063625} (\bibinfo {year} {2012})}\BibitemShut {NoStop}%
\bibitem [{\citenamefont {Hu}\ \emph {et~al.}(2020)\citenamefont {Hu},
  \citenamefont {Wu}, \citenamefont {He}, \citenamefont {Liu},\ and\
  \citenamefont {Hu}}]{Hu2020PhysRevA.101.013615}%
  \BibitemOpen
  \bibfield  {author} {\bibinfo {author} {\bibfnamefont {J.}~\bibnamefont
  {Hu}}, \bibinfo {author} {\bibfnamefont {F.}~\bibnamefont {Wu}}, \bibinfo
  {author} {\bibfnamefont {L.}~\bibnamefont {He}}, \bibinfo {author}
  {\bibfnamefont {X.-J.}\ \bibnamefont {Liu}},\ and\ \bibinfo {author}
  {\bibfnamefont {H.}~\bibnamefont {Hu}},\ }\bibfield  {title} {\bibinfo
  {title} {Theory of strongly paired fermions with arbitrary short-range
  interactions},\ }\href {https://doi.org/10.1103/PhysRevA.101.013615}
  {\bibfield  {journal} {\bibinfo  {journal} {Phys. Rev. A}\ }\textbf {\bibinfo
  {volume} {101}},\ \bibinfo {pages} {013615} (\bibinfo {year}
  {2020})}\BibitemShut {NoStop}%
\bibitem [{\citenamefont {Innocenti}\ \emph {et~al.}(2010)\citenamefont
  {Innocenti}, \citenamefont {Poccia}, \citenamefont {Ricci}, \citenamefont
  {Valletta}, \citenamefont {Caprara}, \citenamefont {Perali},\ and\
  \citenamefont {Bianconi}}]{PhysRevB.82.184528}%
  \BibitemOpen
  \bibfield  {author} {\bibinfo {author} {\bibfnamefont {D.}~\bibnamefont
  {Innocenti}}, \bibinfo {author} {\bibfnamefont {N.}~\bibnamefont {Poccia}},
  \bibinfo {author} {\bibfnamefont {A.}~\bibnamefont {Ricci}}, \bibinfo
  {author} {\bibfnamefont {A.}~\bibnamefont {Valletta}}, \bibinfo {author}
  {\bibfnamefont {S.}~\bibnamefont {Caprara}}, \bibinfo {author} {\bibfnamefont
  {A.}~\bibnamefont {Perali}},\ and\ \bibinfo {author} {\bibfnamefont
  {A.}~\bibnamefont {Bianconi}},\ }\bibfield  {title} {\bibinfo {title}
  {Resonant and crossover phenomena in a multiband superconductor: Tuning the
  chemical potential near a band edge},\ }\href
  {https://doi.org/10.1103/PhysRevB.82.184528} {\bibfield  {journal} {\bibinfo
  {journal} {Phys. Rev. B}\ }\textbf {\bibinfo {volume} {82}},\ \bibinfo
  {pages} {184528} (\bibinfo {year} {2010})}\BibitemShut {NoStop}%
\bibitem [{\citenamefont {Mazziotti}\ \emph {et~al.}(2021)\citenamefont
  {Mazziotti}, \citenamefont {Raimondi}, \citenamefont {Valletta},
  \citenamefont {Campi},\ and\ \citenamefont
  {Bianconi}}]{mazziotti2021resonant}%
  \BibitemOpen
  \bibfield  {author} {\bibinfo {author} {\bibfnamefont {M.~V.}\ \bibnamefont
  {Mazziotti}}, \bibinfo {author} {\bibfnamefont {R.}~\bibnamefont {Raimondi}},
  \bibinfo {author} {\bibfnamefont {A.}~\bibnamefont {Valletta}}, \bibinfo
  {author} {\bibfnamefont {G.}~\bibnamefont {Campi}},\ and\ \bibinfo {author}
  {\bibfnamefont {A.}~\bibnamefont {Bianconi}},\ }\bibfield  {title} {\bibinfo
  {title} {Resonant multi-gap superconductivity at room temperature near a
  lifshitz topological transition in sulfur hydrides},\ }
  {\bibfield  {journal} {\bibinfo  {journal} {Journal of Applied Physics}\
  }\textbf {\bibinfo {volume} {130}},\ \bibinfo {pages} {173904} (\bibinfo
  {year} {2021})}\BibitemShut {NoStop}%
\bibitem [{\citenamefont {Schirotzek}\ \emph {et~al.}(2008)\citenamefont
  {Schirotzek}, \citenamefont {Shin}, \citenamefont {Schunck},\ and\
  \citenamefont {Ketterle}}]{Schirotzek2008PhysRevLett.101.140403}%
  \BibitemOpen
  \bibfield  {author} {\bibinfo {author} {\bibfnamefont {A.}~\bibnamefont
  {Schirotzek}}, \bibinfo {author} {\bibfnamefont {Y.-i.}\ \bibnamefont
  {Shin}}, \bibinfo {author} {\bibfnamefont {C.~H.}\ \bibnamefont {Schunck}},\
  and\ \bibinfo {author} {\bibfnamefont {W.}~\bibnamefont {Ketterle}},\
  }\bibfield  {title} {\bibinfo {title} {Determination of the superfluid gap in
  atomic fermi gases by quasiparticle spectroscopy},\ }\href
  {https://doi.org/10.1103/PhysRevLett.101.140403} {\bibfield  {journal}
  {\bibinfo  {journal} {Phys. Rev. Lett.}\ }\textbf {\bibinfo {volume} {101}},\
  \bibinfo {pages} {140403} (\bibinfo {year} {2008})}\BibitemShut {NoStop}%
\bibitem [{\citenamefont {Stewart}\ \emph {et~al.}(2008)\citenamefont
  {Stewart}, \citenamefont {Gaebler},\ and\ \citenamefont
  {Jin}}]{stewart2008Nature.454.744}%
  \BibitemOpen
  \bibfield  {author} {\bibinfo {author} {\bibfnamefont {J.}~\bibnamefont
  {Stewart}}, \bibinfo {author} {\bibfnamefont {J.}~\bibnamefont {Gaebler}},\
  and\ \bibinfo {author} {\bibfnamefont {D.}~\bibnamefont {Jin}},\ }\bibfield
  {title} {\bibinfo {title} {Using photoemission spectroscopy to probe a
  strongly interacting fermi gas},\ } {\bibfield  {journal}
  {\bibinfo  {journal} {Nature (London)}\ }\textbf {\bibinfo {volume} {454}},\
  \bibinfo {pages} {744} (\bibinfo {year} {2008})}\BibitemShut {NoStop}%
\bibitem [{\citenamefont {Sagi}\ \emph {et~al.}(2015)\citenamefont {Sagi},
  \citenamefont {Drake}, \citenamefont {Paudel}, \citenamefont {Chapurin},\
  and\ \citenamefont {Jin}}]{Sagi2015PhysRevLett.114.075301}%
  \BibitemOpen
  \bibfield  {author} {\bibinfo {author} {\bibfnamefont {Y.}~\bibnamefont
  {Sagi}}, \bibinfo {author} {\bibfnamefont {T.~E.}\ \bibnamefont {Drake}},
  \bibinfo {author} {\bibfnamefont {R.}~\bibnamefont {Paudel}}, \bibinfo
  {author} {\bibfnamefont {R.}~\bibnamefont {Chapurin}},\ and\ \bibinfo
  {author} {\bibfnamefont {D.~S.}\ \bibnamefont {Jin}},\ }\bibfield  {title}
  {\bibinfo {title} {Breakdown of the fermi liquid description for strongly
  interacting fermions},\ }\href
  {https://doi.org/10.1103/PhysRevLett.114.075301} {\bibfield  {journal}
  {\bibinfo  {journal} {Phys. Rev. Lett.}\ }\textbf {\bibinfo {volume} {114}},\
  \bibinfo {pages} {075301} (\bibinfo {year} {2015})}\BibitemShut {NoStop}%
\bibitem [{\citenamefont {Iskin}(2016)}]{Iskin2016PhysRevA.94.011604}%
  \BibitemOpen
  \bibfield  {author} {\bibinfo {author} {\bibfnamefont {M.}~\bibnamefont
  {Iskin}},\ }\bibfield  {title} {\bibinfo {title} {Two-band superfluidity and
  intrinsic josephson effect in alkaline-earth-metal fermi gases across an
  orbital feshbach resonance},\ }\href
  {https://doi.org/10.1103/PhysRevA.94.011604} {\bibfield  {journal} {\bibinfo
  {journal} {Phys. Rev. A}\ }\textbf {\bibinfo {volume} {94}},\ \bibinfo
  {pages} {011604} (\bibinfo {year} {2016})}\BibitemShut {NoStop}%
\bibitem [{\citenamefont {Valtolina}\ \emph {et~al.}(2015)\citenamefont
  {Valtolina}, \citenamefont {Burchianti}, \citenamefont {Amico}, \citenamefont
  {Neri}, \citenamefont {Xhani}, \citenamefont {Seman}, \citenamefont
  {Trombettoni}, \citenamefont {Smerzi}, \citenamefont {Zaccanti},
  \citenamefont {Inguscio} \emph {et~al.}}]{valtolina2015josephson}%
  \BibitemOpen
  \bibfield  {author} {\bibinfo {author} {\bibfnamefont {G.}~\bibnamefont
  {Valtolina}}, \bibinfo {author} {\bibfnamefont {A.}~\bibnamefont
  {Burchianti}}, \bibinfo {author} {\bibfnamefont {A.}~\bibnamefont {Amico}},
  \bibinfo {author} {\bibfnamefont {E.}~\bibnamefont {Neri}}, \bibinfo {author}
  {\bibfnamefont {K.}~\bibnamefont {Xhani}}, \bibinfo {author} {\bibfnamefont
  {J.~A.}\ \bibnamefont {Seman}}, \bibinfo {author} {\bibfnamefont
  {A.}~\bibnamefont {Trombettoni}}, \bibinfo {author} {\bibfnamefont
  {A.}~\bibnamefont {Smerzi}}, \bibinfo {author} {\bibfnamefont
  {M.}~\bibnamefont {Zaccanti}}, \bibinfo {author} {\bibfnamefont
  {M.}~\bibnamefont {Inguscio}}, \emph {et~al.},\ }\bibfield  {title} {\bibinfo
  {title} {Josephson effect in fermionic superfluids across the bec-bcs
  crossover},\ } {\bibfield  {journal} {\bibinfo  {journal}
  {Science}\ }\textbf {\bibinfo {volume} {350}},\ \bibinfo {pages} {1505}
  (\bibinfo {year} {2015})}\BibitemShut {NoStop}%
\bibitem [{\citenamefont {Leggett}(1966)}]{Leggett1966PTP.36.901}%
  \BibitemOpen
  \bibfield  {author} {\bibinfo {author} {\bibfnamefont {A.~J.}\ \bibnamefont
  {Leggett}},\ }\bibfield  {title} {\bibinfo {title} {{Number-Phase
  Fluctuations in Two-Band Superconductors}},\ }\href
  {https://doi.org/10.1143/PTP.36.901} {\bibfield  {journal} {\bibinfo
  {journal} {Prog. Theor. Phys.}\ }\textbf {\bibinfo {volume} {36}},\ \bibinfo
  {pages} {901} (\bibinfo {year} {1966})}\BibitemShut {NoStop}%
\bibitem [{\citenamefont {Lin}\ and\ \citenamefont
  {Hu}(2012)}]{Lin2012PhysRevLett.108.177005}%
  \BibitemOpen
  \bibfield  {author} {\bibinfo {author} {\bibfnamefont {S.-Z.}\ \bibnamefont
  {Lin}}\ and\ \bibinfo {author} {\bibfnamefont {X.}~\bibnamefont {Hu}},\
  }\bibfield  {title} {\bibinfo {title} {Massless leggett mode in three-band
  superconductors with time-reversal-symmetry breaking},\ }\href
  {https://doi.org/10.1103/PhysRevLett.108.177005} {\bibfield  {journal}
  {\bibinfo  {journal} {Phys. Rev. Lett.}\ }\textbf {\bibinfo {volume} {108}},\
  \bibinfo {pages} {177005} (\bibinfo {year} {2012})}\BibitemShut {NoStop}%
\bibitem [{\citenamefont {Hoinka}\ \emph {et~al.}(2017)\citenamefont {Hoinka},
  \citenamefont {Dyke}, \citenamefont {Lingham}, \citenamefont {Kinnunen},
  \citenamefont {Bruun},\ and\ \citenamefont
  {Vale}}]{Hoinka2017Nat.Phys.13.943}%
  \BibitemOpen
  \bibfield  {author} {\bibinfo {author} {\bibfnamefont {S.}~\bibnamefont
  {Hoinka}}, \bibinfo {author} {\bibfnamefont {P.}~\bibnamefont {Dyke}},
  \bibinfo {author} {\bibfnamefont {M.~G.}\ \bibnamefont {Lingham}}, \bibinfo
  {author} {\bibfnamefont {J.~J.}\ \bibnamefont {Kinnunen}}, \bibinfo {author}
  {\bibfnamefont {G.~M.}\ \bibnamefont {Bruun}},\ and\ \bibinfo {author}
  {\bibfnamefont {C.~J.}\ \bibnamefont {Vale}},\ }\bibfield  {title} {\bibinfo
  {title} {Goldstone mode and pair-breaking excitations in atomic fermi
  superfluids},\ } {\bibfield  {journal} {\bibinfo  {journal}
  {Nat. Phys.}\ }\textbf {\bibinfo {volume} {13}},\ \bibinfo {pages} {943}
  (\bibinfo {year} {2017})}\BibitemShut {NoStop}%
\bibitem [{\citenamefont {Behrle}\ \emph {et~al.}(2018)\citenamefont {Behrle},
  \citenamefont {Harrison}, \citenamefont {Kombe}, \citenamefont {Gao},
  \citenamefont {Link}, \citenamefont {Bernier}, \citenamefont {Kollath},\ and\
  \citenamefont {K{\"o}hl}}]{Behrle2018Nat.Phys.14.781}%
  \BibitemOpen
  \bibfield  {author} {\bibinfo {author} {\bibfnamefont {A.}~\bibnamefont
  {Behrle}}, \bibinfo {author} {\bibfnamefont {T.}~\bibnamefont {Harrison}},
  \bibinfo {author} {\bibfnamefont {J.}~\bibnamefont {Kombe}}, \bibinfo
  {author} {\bibfnamefont {K.}~\bibnamefont {Gao}}, \bibinfo {author}
  {\bibfnamefont {M.}~\bibnamefont {Link}}, \bibinfo {author} {\bibfnamefont
  {J.-S.}\ \bibnamefont {Bernier}}, \bibinfo {author} {\bibfnamefont
  {C.}~\bibnamefont {Kollath}},\ and\ \bibinfo {author} {\bibfnamefont
  {M.}~\bibnamefont {K{\"o}hl}},\ }\bibfield  {title} {\bibinfo {title} {Higgs
  mode in a strongly interacting fermionic superfluid},\ }
  {\bibfield  {journal} {\bibinfo  {journal} {Nat. Phys.}\ }\textbf {\bibinfo
  {volume} {14}},\ \bibinfo {pages} {781} (\bibinfo {year} {2018})}\BibitemShut
  {NoStop}%
\bibitem [{\citenamefont {Tajima}\ \emph {et~al.}(2019)\citenamefont {Tajima},
  \citenamefont {Yerin}, \citenamefont {Perali},\ and\ \citenamefont
  {Pieri}}]{Tajima2019PhysRevB.99.180503}%
  \BibitemOpen
  \bibfield  {author} {\bibinfo {author} {\bibfnamefont {H.}~\bibnamefont
  {Tajima}}, \bibinfo {author} {\bibfnamefont {Y.}~\bibnamefont {Yerin}},
  \bibinfo {author} {\bibfnamefont {A.}~\bibnamefont {Perali}},\ and\ \bibinfo
  {author} {\bibfnamefont {P.}~\bibnamefont {Pieri}},\ }\bibfield  {title}
  {\bibinfo {title} {Enhanced critical temperature, pairing fluctuation
  effects, and bcs-bec crossover in a two-band fermi gas},\ }\href
  {https://doi.org/10.1103/PhysRevB.99.180503} {\bibfield  {journal} {\bibinfo
  {journal} {Phys. Rev. B}\ }\textbf {\bibinfo {volume} {99}},\ \bibinfo
  {pages} {180503} (\bibinfo {year} {2019})}\BibitemShut {NoStop}%
\bibitem [{\citenamefont {Salasnich}\ \emph {et~al.}(2019)\citenamefont
  {Salasnich}, \citenamefont {Shanenko}, \citenamefont {Vagov}, \citenamefont
  {Aguiar},\ and\ \citenamefont {Perali}}]{Salasnich2019PhysRevB.100.064510}%
  \BibitemOpen
  \bibfield  {author} {\bibinfo {author} {\bibfnamefont {L.}~\bibnamefont
  {Salasnich}}, \bibinfo {author} {\bibfnamefont {A.~A.}\ \bibnamefont
  {Shanenko}}, \bibinfo {author} {\bibfnamefont {A.}~\bibnamefont {Vagov}},
  \bibinfo {author} {\bibfnamefont {J.~A.}\ \bibnamefont {Aguiar}},\ and\
  \bibinfo {author} {\bibfnamefont {A.}~\bibnamefont {Perali}},\ }\bibfield
  {title} {\bibinfo {title} {Screening of pair fluctuations in superconductors
  with coupled shallow and deep bands: A route to higher-temperature
  superconductivity},\ }\href {https://doi.org/10.1103/PhysRevB.100.064510}
  {\bibfield  {journal} {\bibinfo  {journal} {Phys. Rev. B}\ }\textbf {\bibinfo
  {volume} {100}},\ \bibinfo {pages} {064510} (\bibinfo {year}
  {2019})}\BibitemShut {NoStop}%
\bibitem [{\citenamefont {Shi}\ \emph {et~al.}(2022)\citenamefont {Shi},
  \citenamefont {Zhang},\ and\ \citenamefont {de~Melo}}]{Shi2022}%
  \BibitemOpen
  \bibfield  {author} {\bibinfo {author} {\bibfnamefont {Y.-R.}\ \bibnamefont
  {Shi}}, \bibinfo {author} {\bibfnamefont {W.}~\bibnamefont {Zhang}},\ and\
  \bibinfo {author} {\bibfnamefont {C.~A. R.~S.}\ \bibnamefont {de~Melo}},\
  }\bibfield  {title} {\bibinfo {title} {The evolution from {BCS} to bose
  pairing in two-band superfluids: Quantum phase transitions and crossovers by
  tuning band offset and interactions},\ }\href
  {https://doi.org/10.1209/0295-5075/ac8178} {\bibfield  {journal} {\bibinfo
  {journal} {Europhys. Lett.}\ }\textbf {\bibinfo {volume} {139}},\ \bibinfo
  {pages} {36004} (\bibinfo {year} {2022})}\BibitemShut {NoStop}%
\end{thebibliography}

%

\end{CJK}
\end{document}